\documentclass[10pt]{iopart}
\usepackage{amsfonts}
\usepackage{amssymb}
\usepackage{graphicx}
\usepackage{cite}

\begin{document}

\title{Bose-Einstein Condensate Dark Matter Halos confronted with galactic
rotation curves}
\author{M. Dwornik, Z. Keresztes, E. Kun and L. \'{A}. Gergely \\
Institute of Physics, University of Szeged, D\'{o}m T\'{e}r 9, Szeged 6720,
Hungary}

\begin{abstract}
We present a comparative confrontation of both the Bose-Einstein Condensate (BEC) and the Navarro-Frenk-White (NFW) dark
halo models with galactic rotation curves. We employ 6 High Surface Brightness (HSB), 6 Low Surface Brightness (LSB), and
7 dwarf galaxies with rotation curves falling into two classes. In the first class rotational velocities increase with radius over the
observed range.The BEC and NFW models give comparable fits for HSB and LSB galaxies of this type, while for dwarf galaxies
the fit is significantly better with the BEC model. In the second class the rotational velocity of HSB and LSB galaxies exhibits long
flat plateaus, resulting in better fit with the NFW model for HSB galaxies and comparable fits for LSB galaxies. We conclude that
due to its central density cusp avoidance the BEC model fits better dwarf galaxy dark matter distribution. Nevertheless it suffers
from sharp cutoff in larger galaxies, where the NFW model performs better. The investigated galaxy sample obeys the Tully-Fisher
relation, including the particular characteristics exhibited by dwarf galaxies. In both models the fitting enforces a relation between
dark matter parameters: the characteristic density and the corresponding characteristic distance scale with an inverse power.
\end{abstract}

\section{Introduction}

The visible part of most galaxies is embedded in a dark matter (DM) halo of yet unknown composition, observable only through its gravitational interaction with the baryonic matter. Assuming the standard  $\Lambda $CDM cosmological model, the Planck satellite measurements of the cosmic microwave background anisotropy power spectrum support 4.9\% baryonic matter, 26.8\% DM, and 68.3\% dark energy in the Universe \cite{planck1,planck2}.

Investigation of mass distribution of spiral galaxies is an essential tool in the research of DM. Beside the stellar disk and central bulge, most of the galaxies harbour a spherically symmetric, massive DM halo, which dominates the dynamics in the outer regions of the stellar disk. Nevertheless there are examples of galaxies which at larger radii are better described by a flattened baryonic mass distribution (global disk model) \cite{jaloc}.

Several DM candidates and alternatives have been proposed, the latter assuming Einstein’s theory of gravity breaking down on the galactic scale and above (\cite{milg,sand,mof,mann,roberts,boeh3,boeh1,berto,boeh2}). In brane-world and $f(R)$-gravity models, the galactic rotation curves could be explained without DM (\cite{mak,raha,gerg,stabile}).

At this moment there are strong experimental constraints for all proposed dark matter candidates. Supersymmetric dark matter has been strongly constrained by LHC \cite{supersymmetricLHC1,supersymmetricLHC2}, sterile neutrinos disruled with 99\% confidence level by IceCube \cite{SterileNeutrinoIceCube}, Weekly Interacting Massive Particles (WIMPs) severely bounded by the LUX \cite{Lux}, PandaX-II \cite{Panda2016} and Xenon100 \cite{Xenon100} experiments. Extra dimensional effects as dark matter substitutes have been also contained by LHC \cite{Choudhury}. Massive Compact Halo Objects (MACHOs) with masses below 20 solar masses have been shown to give at most 10\% of dark matter by microlensing experiments on the Large Magellanic Cloud \cite{Macho}. There is still hope for larger mass MACHOs as dark matter candidates, revived after the spectacular first direct detection of gravitational waves \cite{GW}, sourced by black holes of approximately 30 solar masses.

It is well known that hot dark matter (HDM) consisting of light ($m\propto $
eV) particles cannot reproduce the cosmological structure formation, as they
imply that the superclusters of galaxies are the first structures to form
contradicting CMB observations, according to which superclusters would form
at the present epoch \cite{primack}. Warm dark matter ($m\propto $ keV)
models seem to be compatible with the astronomical observations on galactic
and also cosmological scales \cite{vega,weietal}. Leading candidates
for warm dark matter are the right handed neutrinos, which in contrast with
their left handed counterparts do not participate in the weak interaction.
The decay of these sterile neutrinos produces high amount of X-rays, which
can boost the star formation rate leading to an earlier reionization \cite%
{biermann}. The existence of sterile neutrinos was however severely
constrained by recent IceCube Neutrino Observatory experiments \cite{IceCUBE}%
. Cold dark matter (CDM) also shows remarkably good agreement with
observations over kpc scales (\cite{padm,peeb}). Particular CDM
candidates, like neutralinos (which are stable and can be produced thermally
in the early Universe) and other WIMPs originating in supersymmetric extensions of the Standard Model were
severely constrained by recent LHC results, rendering them into the range $%
200GeV\lesssim m_{n}\lesssim 500GeV$ \cite{fowlie}. In a Higgs-portal DM
scenario the Higgs boson acts as the mediator particle between DM and
Standard Model particles, and it can decay to a pair of DM particles. Very
recent constraints established by the ATLAS Collaboration on DM-nucleon
scattering cross section impose upper limits of approximately 60 GeV for
each of the scalar, fermion and vector DM candidates (see Fig. 4 of Ref. 
\cite{atlas}), within the framework of this scenario. While MACHOs of masses less then 10 solar masses (like white
dwarfs, neutron stars, brown dwarfs and unassociated planets, primordial
black holes in the astrophysical mass range) were disruled either by Big
Bang Nucleosynthesis constraints or microlensing experiments as dominant DM
candidates, primordial black holes with intermediate mass could still be
viable candidates \cite{Frampton,Sasaki}.

Large N-body simulations (e.g. \cite{millennium}) performed in the framework
of the $\Lambda $CDM-model ($\Lambda $ being the cosmological constant) were
compatible with CDM halos with central density cusps \cite{nav}. They are
modeled by the Navarro-Frenk-White (NFW) DM\ density profile $\rho
_{NFW}(r)=\rho _{s}/(r/r_{s})(1+r/r_{s})^{2}$, where $r_{s}$ is a scale
radius and $\rho _{s}$ is a characteristic density. Some observations
support such a steep cuspy density profile \cite{valenzuela,jardel},
nevertheless certain high-resolution rotation curves instead indicate that
the distribution of DM in the centres of DM dominated dwarf and Low Surface
Brightness (LSB) galaxies is much shallower, exhibiting a core with nearly
constant density \cite{burkert}. In turn, the baryonic matter distribution
may also affect the DM density profile. As shown in \cite{teyssier} a
dark matter core within an isolated, initially cuspy dark matter halo may
form due to strong supernova feedback. By contrast, adiabatic contraction of
baryonic gas tends to produce even cuspier dark matter halos \cite{inoue}.

The surface number-density profiles of satellites decline with the projected
distance as a power law with the slope $\left( -2\right) \div \left(
-1.5\right) $, while the line-of-sight velocity dispersions decline
gradually \cite{klypin2009}. These observations support the NFW model on
scales of $50\div 500$ kpc.

In a cosmological setup various scalar field DM models were also discussed (%
\cite{gt,rodr} and references therein). A particular scalar field DM
model describes light bosons in a dilute gas. The thermal de Broglie
wavelength of the particles is $\lambda _{T}\propto 1/\sqrt{mT}$, which can
be large for light bosons ($m<$eV) and for low temperature. Below a critical
temperature ($T_{c}$), the bosons' wave packets, which are the order of $%
\lambda _{T}$ overlap, resulting in correlated particles. Such bosons share
the same quantum ground state, behaving as a Bose-Einstein condensate (BEC),
characterized by a single macroscopic wave function. It has been proposed
that galactic DM halos could be gigantic BECs \cite{sin}.

It has been shown that caustics of ring shape appear in rotating BEC models, which have an effect on rotation curves, by causing bumps \cite{Sikivie1, Sikivie2}. Such ring shaped caustics degenerate into the origin in the non-rotating BEC limit, adopted in this paper.

 The self gravitating condensate is described by the Gross-Pitaevskii-Poisson equation
system in the mean-field approximation \cite{Gross1}, \cite{Gross2}, \cite%
{Pitaevskii}, \cite{boeh1}. In the Thomas-Fermi approximation, a 2-parameter
(mass $m$ and scattering length $a$) density distribution of the BEC halo is
obtained [see Eq. (\ref{dens}) below],$~$which is less concentrated towards
the centre as compared to the NFW model, relaxing the cuspy halo problem.

In model \cite{harko11} where a normal dark matter phase with an equation of
state $P=\rho c^{2}\sigma _{tr}^{2}$ condensed into a BEC with
self-interaction ($\sigma _{tr}=0.0017$ being the one-dimensional velocity
dispersion and $c$ the speed of light), the stability of the BEC halo
depends on the particle mass and scattering length. For a given mass the
stability occurs for larger scattering length and for given scattering
length the stability appears at smaller mass. For the scattering lengths: $%
a=10^{3}$ fm, $a=10^{-14}$ fm and $a=10^{-55}$ fm the mass of the BEC
particle arises as $m>1$ eV, $m>2\times 10^{-6}$ eV and $m>4.57\times
10^{-20}$ eV, respectively. Galactic size stable halos can form with $%
m>10^{-24}$ eV (Fig. 3 in Ref. \cite{souza14}). 

A stable BEC halo can form as a result of gravitational collapse \cite%
{harko14}. The model has been tested on kpc scales confronting it with
galactic rotation curve observations \cite{boeh1}. It was pointed out by 
\cite{velten} that the effects of BEC DM should be seen in the matter power
spectrum if the boson mass is in the range $15$ meV $<m<35$ meV and $300$
meV $<m<700$ meV for the scattering lengths $a=10^{6}$ fm and $a=10^{10}$
fm, respectively. In Ref. \cite{lee} the authors showed that the observed
collisional behaviour of DM in the Abell 520 cluster can also be recovered
within the framework of the BEC model. All of the mentioned BEC particle
masses are consistent with the limit $m<1.87$ eV imposed from galaxy
observations and N-body simulation \cite{boyan}. A discrepancy was however
pointed out between the best fit density profile parameters derived from the
strong lensing and the galactic rotational curves data. As a conclusion the
BEC halo should be denser in lens galaxies than in dwarf spheroidals \cite%
{alma}.

In this work we critically examine the BEC model as a possible DM candidate
against rotation curve data, pointing out both advantages and disadvantages
over the NFW model. Previous studies on the compatibility of the BEC\ model
and galactic rotation curves were promising, but relied on a less numerous
and less diversified set of galaxies then employed here (\cite{robl}, \cite%
{dwor}). The paper has the following structure. The basic properties of the BEC
DM model are reviewed in Section \ref{section2}. In Section \ref{section3} a comparison is made between the theoretical predictions
of the BEC model and the observed rotation curve data of three
types of galaxies, the High Surface Brightness (HSB), LSB and dwarf
galaxies. The conclusions are presented in Section \ref{section4}.

\section{The Bose-Einstein condensate galactic dark matter halo}
\label{section2}
An ideal, dilute Bose gas at very low temperature forms a Bose-Einstein
condensate in which all particles are in the same ground state. In the
thermodynamic limit, the critical temperature for the condensation is $%
T_{c}=2\pi \hbar ^{2}\left( n/\zeta \right) ^{3/2}/mk_{B}$ \cite{pita}. Here 
$n$ and $m$ are the number density and the mass of the bosons, respectively, 
$\zeta =2.612$ is a constant, while $\hbar $ and $k_{B}$ denote the reduced
Planck and Boltzmann constants, respectively. Atoms can be regarded as
quantum-mechanical wave packets of the order of their thermal de Broglie
wavelength $\lambda _{T}=\sqrt{2\pi \hbar ^{2}/\left( mk_{B}T\right) }$. The
condition for the condensation $T<T_{c} $ can be reformulated as $l<\lambda
_{T}/\zeta ^{-1/3}$, where $l$ is the average distance between pairs of
bosons, and it occurs when the temperature, hence the
momentum of the bosons, decreases and as a consequence their de Broglie wavelengths overlap.
The thermodynamic
limit is only approximately realized, the finite size giving corrections to
the critical temperature \cite{GrossmannHolthaus,KetterleDruten,KristenToms,Haugerudetal}. A dilute, non-ideal Bose gas also
displays BEC, on the other hand, the condensate fraction is smaller than unity at zero
temperature and the critical temperature is also modified \cite{Giorginietal,Glaumetal,Schutte,dalfovo}. Experimentally, BEC
(which could be formed by bosonic atoms, but also form fermionic Cooper
pairs) has been realized first in $^{87}$Rb \cite{anderson,Han98,Ernst98}, then in $^{23}$Na \cite{Davis95,Hau98}, and in $^{7}$%
Li \cite{Bradley95}.

In a dilute gas, only two-particle interactions dominate. The repulsive,
two-body interparticle potential is approximated as $V_{self}=\lambda \delta
\left( \mathbf{r}-\mathbf{r}^{\prime }\right) $, with a self-coupling
constant $\lambda =4\pi \hbar ^{2}a/m$, where $a$ is the scattering length.
Then in the mean-field approximation (in case when we neglect the contribution of the
excited states) the BEC is described by the Gross-Pitaevskii equation \cite%
{Gross1,Gross2,Pitaevskii}: 
\begin{equation}
i\hbar \frac{\partial }{\partial t}\psi (\mathbf{r},t)=\left[ -\frac{\hbar
^{2}}{2m}\Delta +V_{selfgrav}\left( \mathbf{r}\right) +\lambda \rho \left( 
\mathbf{r},t\right) \right] \psi (\mathbf{r},t)~,  \label{Gross-Pitaevskii}
\end{equation}%
where $\psi (\mathbf{r},t)$ is the wave function of the condensate and $%
\Delta $ is the 3-dimensional Laplacian. The probability density $\rho
\left( \mathbf{r},t\right) =\left\vert \psi (\mathbf{r},t)\right\vert ^{2}$
is normalized to 
\begin{equation}
n_{0}\left( t\right) =\int d\mathbf{r}\rho \left( \mathbf{r},t\right) ~,
\label{norm}
\end{equation}%
where $n_{0}\left( t\right) $ is the number of particles in the ground state
and $\rho \left( \mathbf{r},t\right) $ the number density of the condensate.
The potential $V_{selfgrav}\left( \mathbf{r}\right) /m$ is the Newtonian
gravitational potential produced by the Bose-Einstein condensate.

Stationary solutions of the Gross-Pitaevskii equation can be found in a
simple way by using the Madelung representation of complex wave-functions 
\cite{Madelung,Sonego}, then deriving the Madelung hydrodynamic
equations \cite{Madelung}. Madelung's equations can be interpreted as the
continuity and Euler equations of fluid mechanics, with quantum corrections
included. However, the quantum correction potential in the generalized Euler
equation contributes significantly only close to the boundary of the system 
\cite{Wang}. In the Thomas-Fermi approximation the quantum correction
potential is neglected compared to the self-interaction term. This
approximation becomes more accurate as the particle number increases \cite{Liebetal}.

Assuming a spherically symmetric distribution of the condensate the
following solution was found \cite{Wang,boeh1}: 
\begin{equation}
\rho _{BEC}\left( r\right) =\rho _{BEC}^{\left( c\right) }\frac{\sin kr}{kr}
~,  \label{dens}
\end{equation}
where $\rho _{BEC}=m\rho \left( r\right) $ and 
\begin{equation}
k=\sqrt{\frac{Gm^{3}}{\hbar ^{2}a}}~.
\end{equation}
The central density $\rho _{BEC}^{\left( c\right) }\equiv \rho _{BEC}\left(
0\right) $ is determined from the normalization condition (\ref{norm}) as 
\begin{equation}
\rho _{BEC}^{\left( c\right) }=\frac{n_{0}mk^{3}}{4\pi ^{2}}~.
\end{equation}
The Thomas-Fermi approximation remains valid for $n_{0}\gg 1/ka$ \cite{Wang}.

The BEC galactic DM halo's size is defined by $\rho (R_{BEC})=0$,
giving $k=\pi /R_{BEC}$, i.e. 
\begin{equation}
R_{BEC}=\pi \sqrt{\frac{\hbar ^{2}a}{Gm^{3}}}~.
\end{equation}
The mass profile of the BEC\ halo is then given as 
\begin{eqnarray}
m_{BEC}(r) &=&4\pi \int_{0}^{r}\rho _{BEC}(r)r^{2}dr  \nonumber \\
&=&\frac{4\pi \rho _{BEC}^{(c)}}{k^{2}}r\left( \frac{\sin kr}{kr}-\cos
kr\right) ~.
\end{eqnarray}
The BEC halo contributes to the velocity profile of the particles which are moving
on circular orbit as dictated by the Newtonian gravitational force \cite%
{boeh1}. This can be taken into account by the following equation:
\begin{equation}
v^{2}\left( r\right) =\frac{4\pi G\rho _{BEC}^{(c)}}{k^{2}}\left( \frac{\sin
kr}{kr}-\cos kr\right) ~,  \label{vel}
\end{equation}
which needs to be added to the baryonic contribution respectively.

\section{Confronting the model with rotation curve data}
\label{section3}
The validity of our model was tested by confronting the rotation curve
data of a sample of 6 HSB, 6 LSB and 7 dwarf galaxies, with both the NFW DM
and the BEC density profiles. For reasons to become obvious during our
analysis, we split both the HSB and LSB data sets into two groups (type I.
and II.), based on the shapes of the curves. In the first group the
rotational velocities increase over the whole observed range, while in the
second set the rotation curves exhibit long flat regions.

The commonly used NFW model has the mass density profile

\begin{equation}
\rho _{NFW}(r)=\frac{\rho _{s}}{\left( r/r_{s}\right) \left(
1+r/r_{s}\right) ^{2}}~,
\end{equation}
where $\rho _{s}$ and $r_{s}$ are a characteristic density and distance
scale, to be determined from the fit.

The mass within a sphere with radius $r=yr_{s}$ is then given as 
\begin{equation}
m_{NFW}(r)=4\pi \rho _{s}r_{s}^{3}\left[ \ln (1+y)-\frac{y}{1+y}\right]
\end{equation}
where $y$ is a positive dimensionless radial coordinate.

\subsection{HSB galaxies}

\begin{figure*}[tbp]
\begin{center}
\begin{tabular}{ccc}
\includegraphics[height=5cm, angle=270]{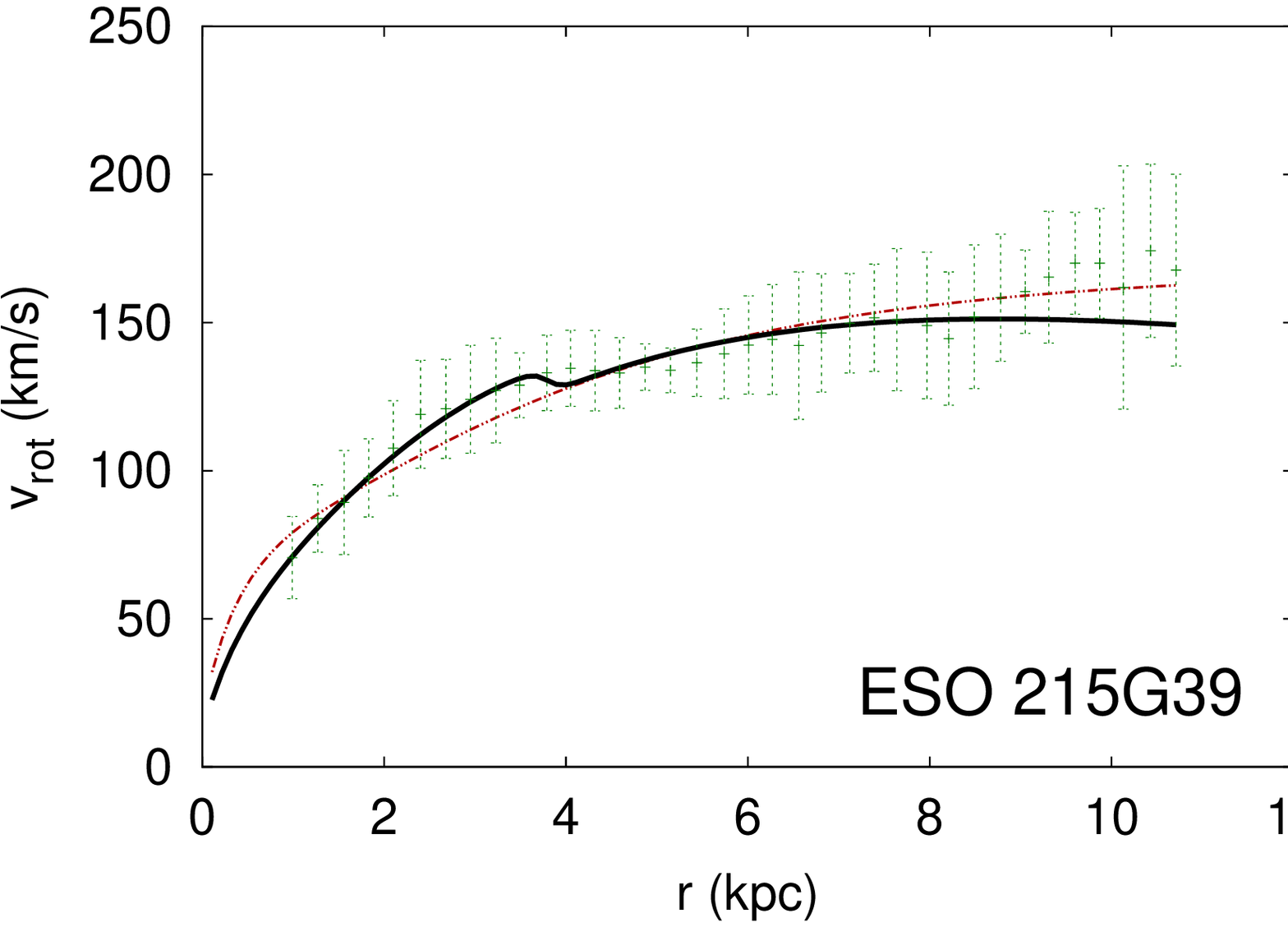} & %
\includegraphics[height=5cm, angle=270]{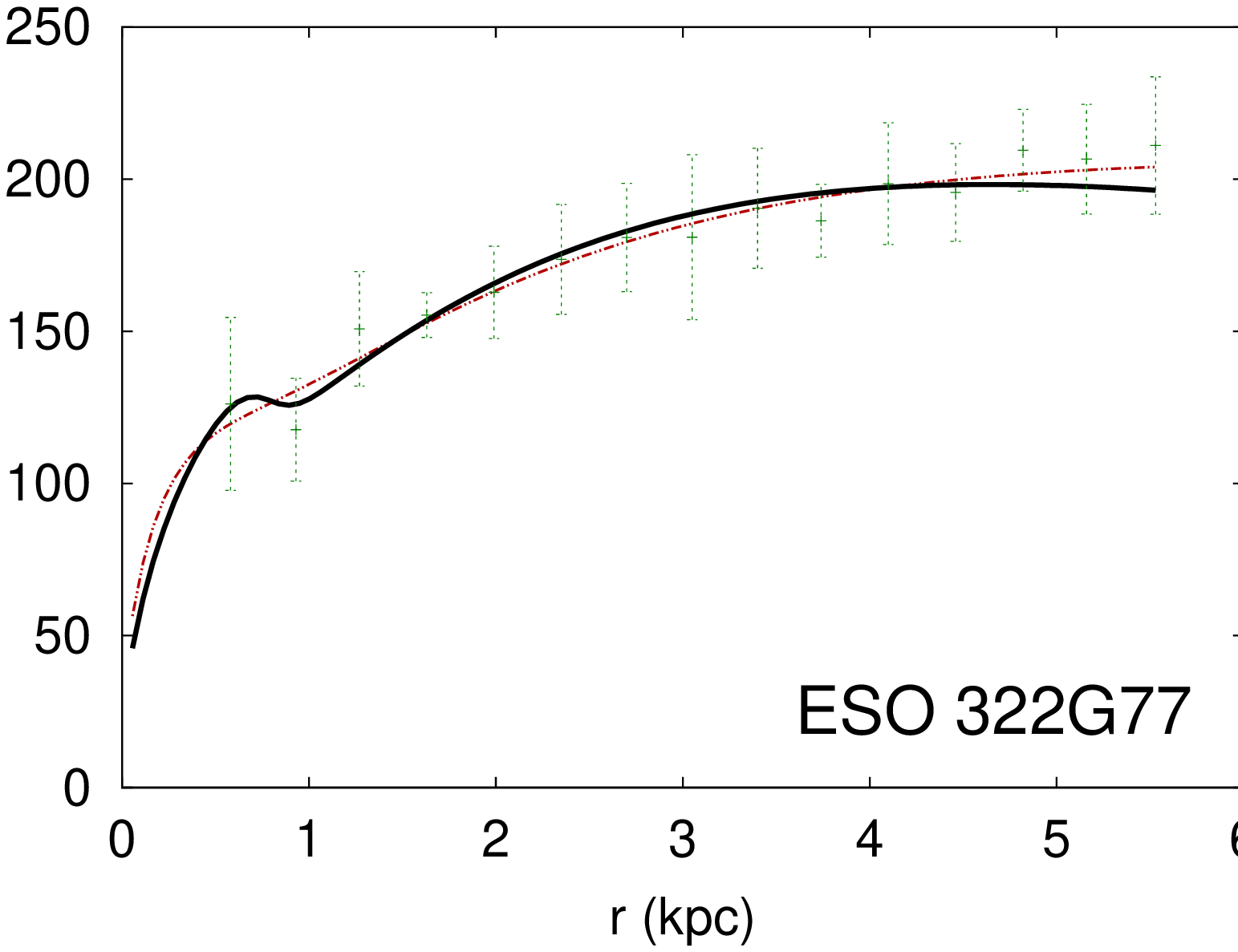} & %
\includegraphics[height=5cm, angle=270]{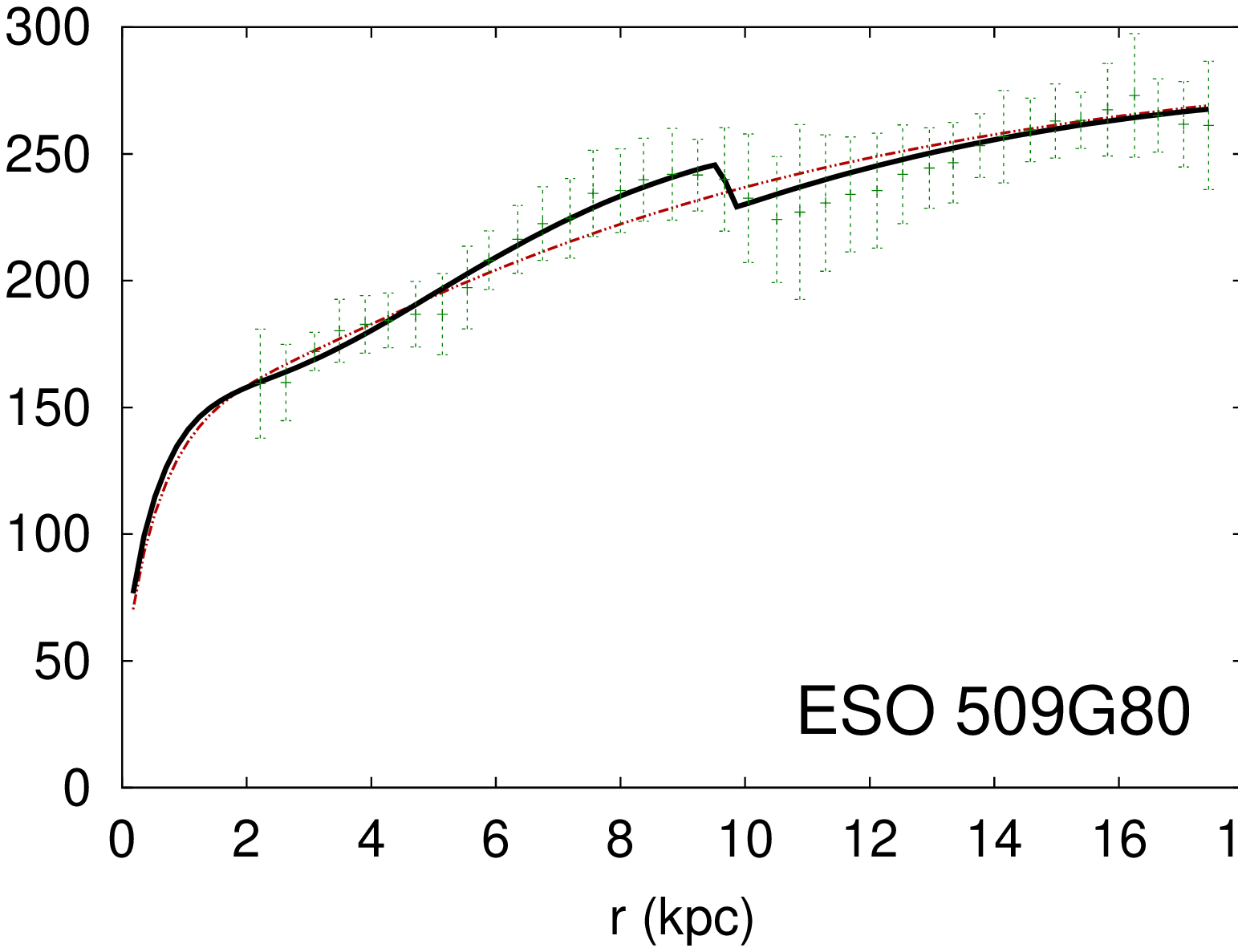} \\ 
&  & 
\end{tabular}%
\end{center}
\caption{Best fit curves for the HSB I. galaxy sample where the solid black
lines hold for the baryonic matter + BEC model, while the dashed red lines
refer to the baryonic matter + NFW model.}
\label{rotcurvHSB1}
\end{figure*}

\begin{figure*}[tbp]
\begin{center}
\begin{tabular}{ccc}
\includegraphics[height=5cm, angle=270]{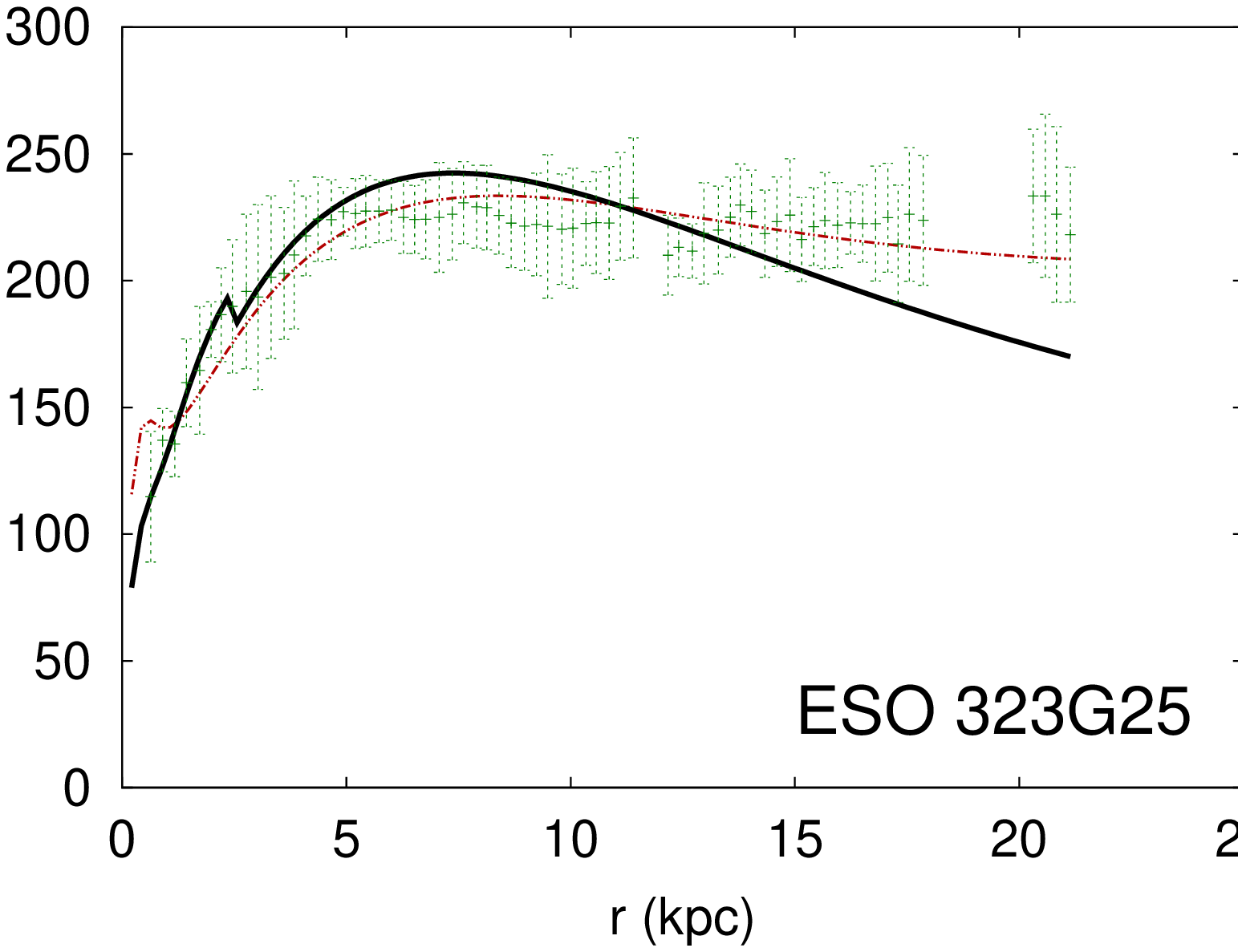} & %
\includegraphics[height=5cm, angle=270]{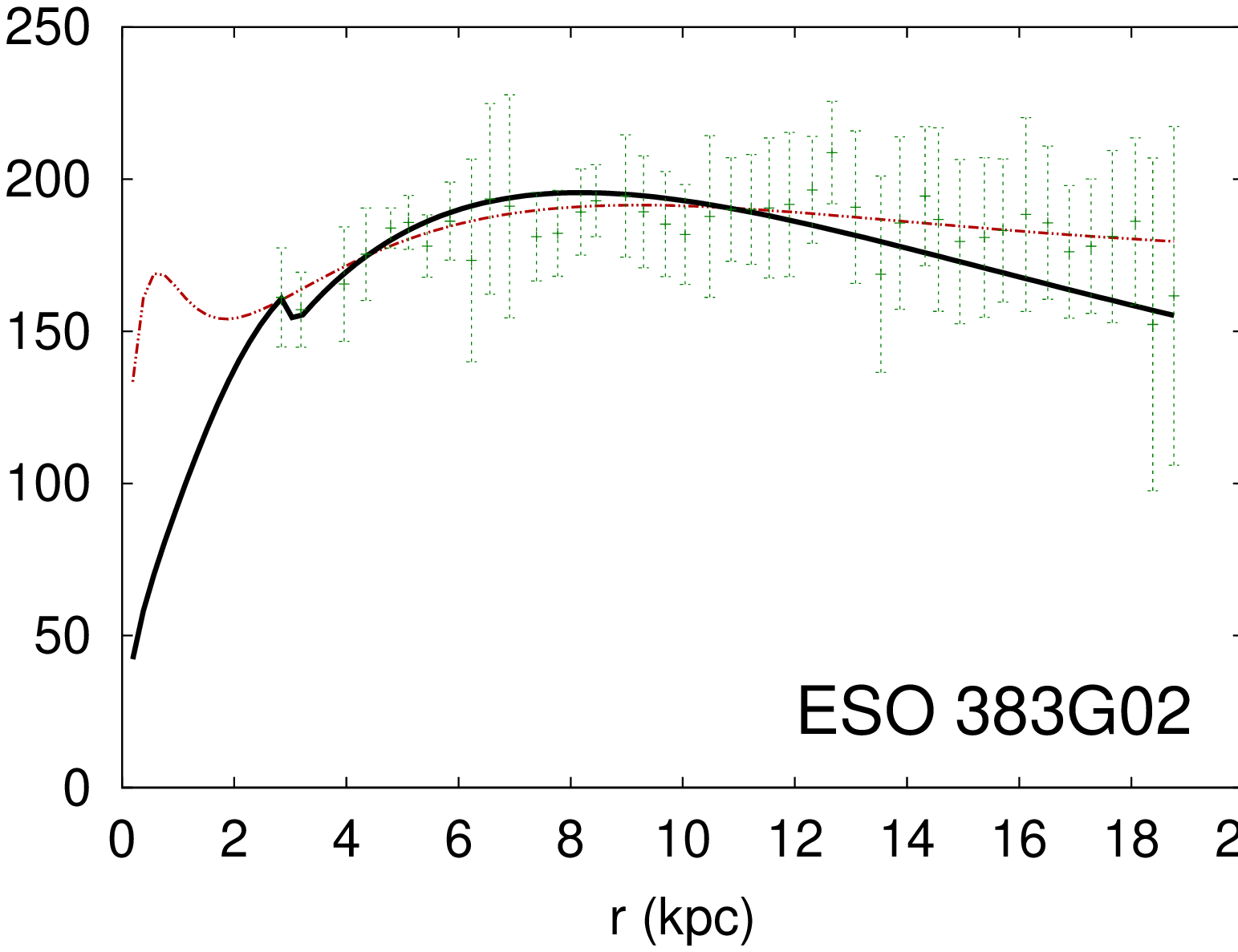} & %
\includegraphics[height=5cm, angle=270]{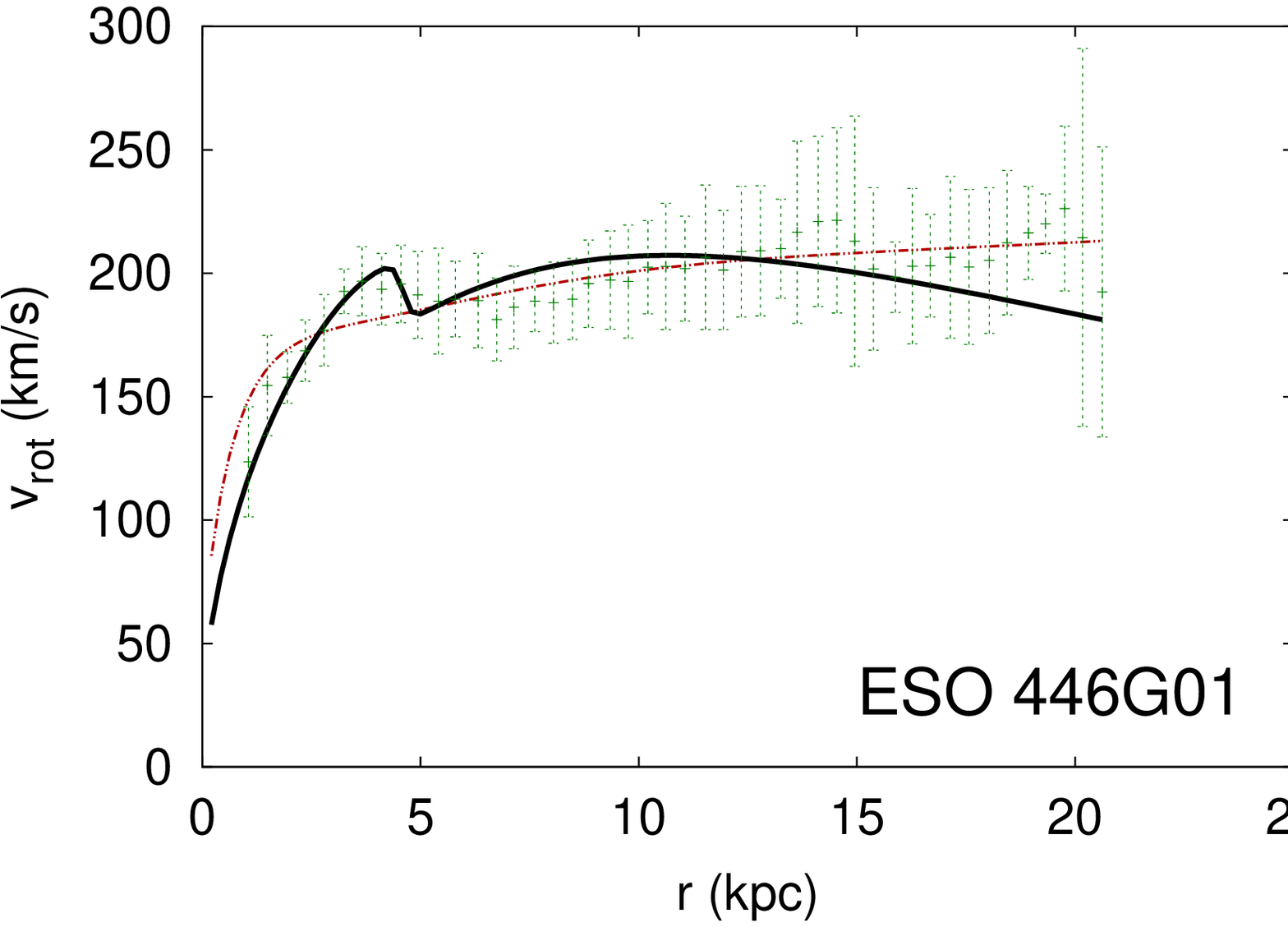} \\ 
&  & 
\end{tabular}%
\end{center}
\caption{Best fit curves for the HSB II. galaxy sample. The solid black
lines hold for the baryonic matter + BEC model, while the dashed red lines
for the baryonic matter + NFW model. The BEC model does not describe well
the extended flat regions.}
\label{rotcurvHSB2}
\end{figure*}

\begin{table*}[tbp]
\begin{center}
\begin{tabular}{c|c|c|c|c|c|c|c}
Galaxy & $D$ & $I_{0,b}$ & $n$ & $r_{0}$ & $r_{b}$ & $I_{0,d}^{HSB}$ & $%
h^{HSB}$ \\ \hline
& Mpc & $\mathrm{mJy/arcsec} ^2$ &  & kpc & kpc & $\mathrm{mJy/arcsec} ^2$ & 
kpc \\ \hline\hline
ESO215G39 & 61.29 & 0.1171 & 0.6609 & 0.78 & 2.58 & 0.0339 & 4.11 \\ 
ESO322G77 & 38.19 & 0.1949 & 0.7552 & 0.33 & 1.37 & 0.0744 & 2.20 \\ 
ESO509G80 & 92.86 & 0.2090 & 0.7621 & 1.10 & 4.69 & 0.0176 & 11.03 \\ 
\hline\hline
ESO323G25 & 59.76 & 0.1113 & 0.4626 & 0.43 & 0.99 & 0.0825 & 3.47 \\ 
ESO383G02 & 85.40 & 0.6479 & 0.7408 & 0.42 & 1.94 & 0.5118 & 3.82 \\ 
ESO446G01 & 98.34 & 0.2093 & 0.8427 & 1.28 & 6.33 & 0.0357 & 5.25 \\ 
&  &  &  &  &  &  & 
\end{tabular}%
\end{center}
\caption{The distances ($D$) and the photometric parameters of the 6 HSB
galaxy sample as determined by the fit with available photometric
data \protect\cite{palu}. Bulge parameters: the central surface brightness (%
$I_{0,b}$), the shape parameter ($n$), the characteristic radius ($r_{0}$)
and radius of the bulge ($r_{b}$). Disk parameters: central surface
brightness ($I_{0,d}^{HSB}$) and length scale ($h^{HSB}$) of the disk.}
\label{Tablehsbphoto}
\end{table*}

In this subsection we will follow the method described in \cite{gerg}. In a
HSB galaxy the baryonic component was decomposed into a thin stellar disk and
a spherically symmetric bulge. It was assumed that the mass distribution of bulge
component follows the de-projected luminosity distribution with a factor
known as the mass-to-light ratio. The bulge parameters were estimated from a S%
\'{e}rsic $r^{1/n}$ bulge model, which was obtained by the fitting of the optical I-band galaxy light
profiles.

Each galaxy's spheroidal bulge component has a surface 
brightness profile which is described by a generalized S\'{e}rsic function \cite{ser} 
\begin{equation}
I_{b}(r)=I_{0,b}\exp \left[ -\left( \frac{r}{r_{0}}\right) ^{1/n}\right] ,
\label{Ib}
\end{equation}
wherein $I_{0,b}$ is the central surface brightness of the bulge, $r_{0}$ is
the characteristic radius of the bulge and the magnitude-radius curve's shape parameter is denoted by $n$.

The mass-to-light ratio for the Sun is
$\gamma _{\odot }=5133$ kg W$^{-1}$. The mass-to-light ratio of the bulge $%
\sigma $ will be given in units of $\gamma _{\odot }$ (solar units). We will
also give the mass in units of the solar mass $M_{\odot }=1.98892\times
10^{30}$ kg. We assume that the radial distribution of visible mass follows 
the radial distribution of light derived from the bulge-disk
decomposition. Accordingly the mass of the bulge inside the projected radius $r$ can be derived from
 the surface brightness observed within this radius: 
\[
m_{b}(r)=\sigma \frac{\mathcal{N}(D)}{F_{\odot }}2\pi
\int\limits_{0}^{r}I_{b}(r)rdr, 
\]
where $F_{\odot }\left( D\right) $ is the apparent flux density of the Sun
at a distance $D$ Mpc, $F_{\odot }\left( D\right) =2.635\times
10^{6-0.4f_{\odot }}\;\mathrm{mJy}{\ }$, with $f_{\odot }=4.08+5\lg \left(
D/1\;\mathrm{Mpc}\right) +25~\mathrm{mag}$, and 
\begin{equation}
\mathcal{N}(D)=4.4684\times 10^{-35}D^{-2}\;\mathrm{m}^{-2}\;\mathrm{arcsec}
^{2}.
\end{equation}
The rotational velocity related to the bulge
\begin{equation}
v_{b}^{2}(r)=\frac{Gm_{b}(r)}{r},
\end{equation}
where $G$ is the gravitational constant.

In case of a spiral galaxy, the radial surface brightness profile of the disk,
decreases exponentially as a function of the radius \cite{free} 
\begin{equation}
I_{d}(r)=I_{0,d}^{HSB}\exp \left( -\frac{r}{h^{HSB}}\right) ,  \label{Id}
\end{equation}
where $I_{0,d}^{HSB}$ is the central surface brightness of the disk and $h^{HSB}$
is a characteristic disk length scale. The disk contributes to the
circular velocity as follows (\cite{free}) 
\begin{equation}
v_{d}^{2}(x)=\frac{GM_{D}^{HSB}}{2h^{HSB}}x^{2}(I_{0}K_{0}-I_{1}K_{1}) ,
\end{equation}
where $x=r/h^{HSB}$ and $I_{n}$ and $K_{n}$ are the modified Bessel
functions evaluated at $x/2$, while $M_{D}^{HSB}$ is the total mass of the
disk.

Accordingly in a HSB galaxy the rotational velocity adds up as 
\begin{eqnarray}
v_{tg}^{2}(x)&=&v_{b}^{2}(x)+v_{d}^{2}(x)+v^2_{DM}(x) ~.  \label{HSBrot}
\end{eqnarray}

In order to validate the BEC+baryonic model, we confront it with rotation curve data of 6 well-tested galaxies (which were already employed in \cite{gerg} for testing a brane-world model). The data was obtained from a sample given in \cite{palu}, and meets the following criteria: (i) it has to be among the best accuracies obtained from the sample and (ii) the bulge has to be spherically symmetric. As a check we also fitted the NFW + baryonic model with the same data set.
The
respective rotation curves are plotted for both models on Figs. \ref%
{rotcurvHSB1} and \ref{rotcurvHSB2}. The small humps on both figures are due
to the baryonic component. From the available photometric data the best fitting
values were derived for the baryonic model parameters  $I_{0,b}$, $n$, $r_{0}$%
, $r_{b}$, $I_{0,d}^{HSB}$ $h^{HSB}$.
By fitting BEC and NFW models to the investigated rotation curve data, the parameters for these 
models (as well as the corresponding baryonic parameters) 
were calculated. The parameter values are indicated in Tables~\ref{Tablehsbphoto} and \ref%
{Tablehsbpar}.

Both the BEC and NFW DM models give comparable $\chi _{\min }^{2}$ values
(within 1$\sigma $\ confidence level) for HSB I galaxies. In case of
galaxies with extended flat regions (HSB II), the NFW DM model fits better
the rotation curves, nevertheless BEC model give rotational curves which
fall outside the 1$\sigma $ confidence level.

\begin{table*}[tbp]
\begin{center}
\resizebox{16.5cm}{!} {
\begin{tabular}{c|c|c|c|c|c|c|c|c|c|c|c}
Galaxy & $\sigma $(BEC) & $M_{D}^{HSB}$(BEC) & $R_{BEC}$ & $\rho^{(c)}_{BEC}$ & $\chi _{\min }^{2}$ (BEC)&
$\sigma $(NFW)& $M_{D}^{HSB}$(NFW) & $r_{s}$ & $\rho_{s}$ & $\chi _{\min }^{2}$(NFW) &1$\sigma$ \\ \hline
 & $\odot$ & $10^{10} M_{\odot} $ & $kpc$ & $10^{-21}kg/m^3$ &   & $\odot$ & $10^{10} M_{\odot}$ & $kpc$ & $10^{-24}kg/m^3$ & \\ \hline\hline
ESO215G39 & 0.3 & 5.61 & 3.8 & 2.0 & 23.07 & 0.6 & 3.84 & 187 & 14.7 & 22.22 & 34.18 \\
ESO322G77 & 1.6 & 5.1 & 0.8 & 89.0 & 9.15 & 2.5 & 3.79 & 709 & 8 & 7.69 & 11.53\\
ESO509G80 & 1.4 & 48.74 & 9.7 & 1.2 & 12.52 & 0.9 & 11 & 22 & 800 & 33.48 & 36.3\\ \hline\hline
ESO323G25 & 2.5 & 12.18 & 2.5 & 11.8 & 222.74 & 6 & 9.43 & 436 & 6 & 80.55 & 66.74\\
ESO383G02 & 0.13 & 8.77 & 3.0 & 5.7 & 48.83 & 1.7 & 6.32 & 459 & 4.2 & 23.3 & 47.9 \\
ESO446G01 & 0.6 & 12.77 & 4.6 & 5.9 & 86.02 & 1.4 & 6.7 & 786 & 4.1 & 43.37 & 44.74 \\
\end{tabular}
}
\end{center}
\caption{The best fit parameters and the minimum values ($\protect\chi %
_{\min }^{2}$) of the $\protect\chi ^{2}$ statistics for the HSB I and II
galaxies (the first and last three galaxies, respectively). Columns 2-5 give
the BEC model parameters (radius $R_{BEC}$ and central density $\protect\rho %
_{BEC}^{(c)}$ of the BEC halo) and the corresponding baryonic parameters
(mass-to-light ratio $\protect\sigma \left( BEC\right) $ of the bulge and
total mass of the disk $M_{D}^{HSB}\left( BEC\right) $). Columns 7-10 give
the NFW model parameters (scale radius $r_{s}$ and characteristic density $%
\protect\rho _{s}$ of the halo) and the corresponding baryonic parameters
(mass-to-light ratio $\protect\sigma \left( NFW\right) $ of the bulge and
total mass of the disk $M_{D}^{HSB}\left( NFW\right) $). The 1$\protect%
\sigma $ confidence levels are shown in the last column (these are the same
for both models). For HSB I galaxies the two models give similar $\protect%
\chi _{\min }^{2}$ values (within 1$\protect\sigma $ confidence level),
however in case of HSB II galaxies with extended flat regions, the NFW model
fits better the rotation curves. The $\protect\chi _{\min }^{2}$ values in
the case of BEC model are outside the 1$\protect\sigma $ confidence level
for HSB II galaxies.}
\label{Tablehsbpar}
\end{table*}

\subsection{LSB galaxies}

The surface brightness of LSB galaxies is substantially fainter than the brightness of the sky at 
night. They belong to an earyl stage class of
galaxies \cite{imp}. LSB galaxies were found to be metal poor, which indicates
a lower star formation rate than what is generaly found in HSB galaxies \cite%
{mcg}. Wide spectrum of colors can be measured in case of LSB galaxies ranging from red to blue \cite%
{neil} and they are diverse as regards morphologies and other properties. 
Most of the LSB galaxies that were observed are dwarf galaxies, however there is also a
significant number of large spirals among LSB galaxies \cite{beij}.

According to our model the LSB galaxy is made up of two main components; one being a 
thin stellar+gas disk and the other one being a CDM component
which is assumed to be a BEC. We use the same model for the disk component as in the case of 
the HSB galaxies. The surface brightness profile can be described by the following equation \cite{free} 
\[
I_{d}(r)=I_{0,d}^{LSB}\exp \left( -\frac{r}{h^{LSB}}\right) ~, 
\]%
where $I_{0,d}^{LSB}$ is the central surface brightness and $h^{LSB}$ the
disk length scale. The contribution of the disk to the circular
velocity can be expressed as 
\begin{equation}
v_{d}^{2}(r)=\frac{GM_{D}^{LSB}}{2h^{LSB}}q^{2}(I_{0}K_{0}-I_{1}K_{1}),
\end{equation}%
where $q=r/h^{LSB}$ and $M_{D}^{LSB}$ is the total mass of the disk while
the modified Bessel functions $I_{n}$ and $K_{n}$ are evaluated at $q/2$.

Consequently, for an arbitrary projected radius $r$ the rotational velocity can be calculated
based on the combined model resulting in the following equation
\[
v_{tg}^{2}(r)=v_{d}^{2}(r)+v_{DM}^{2}~. 
\]

A preliminary check confirmed that the BEC+baryonic model represents a
better fit than the purely BEC model.

\begin{figure*}[tbp]
\begin{center}
\begin{tabular}{ccc}
\includegraphics[height=5cm, angle=270]{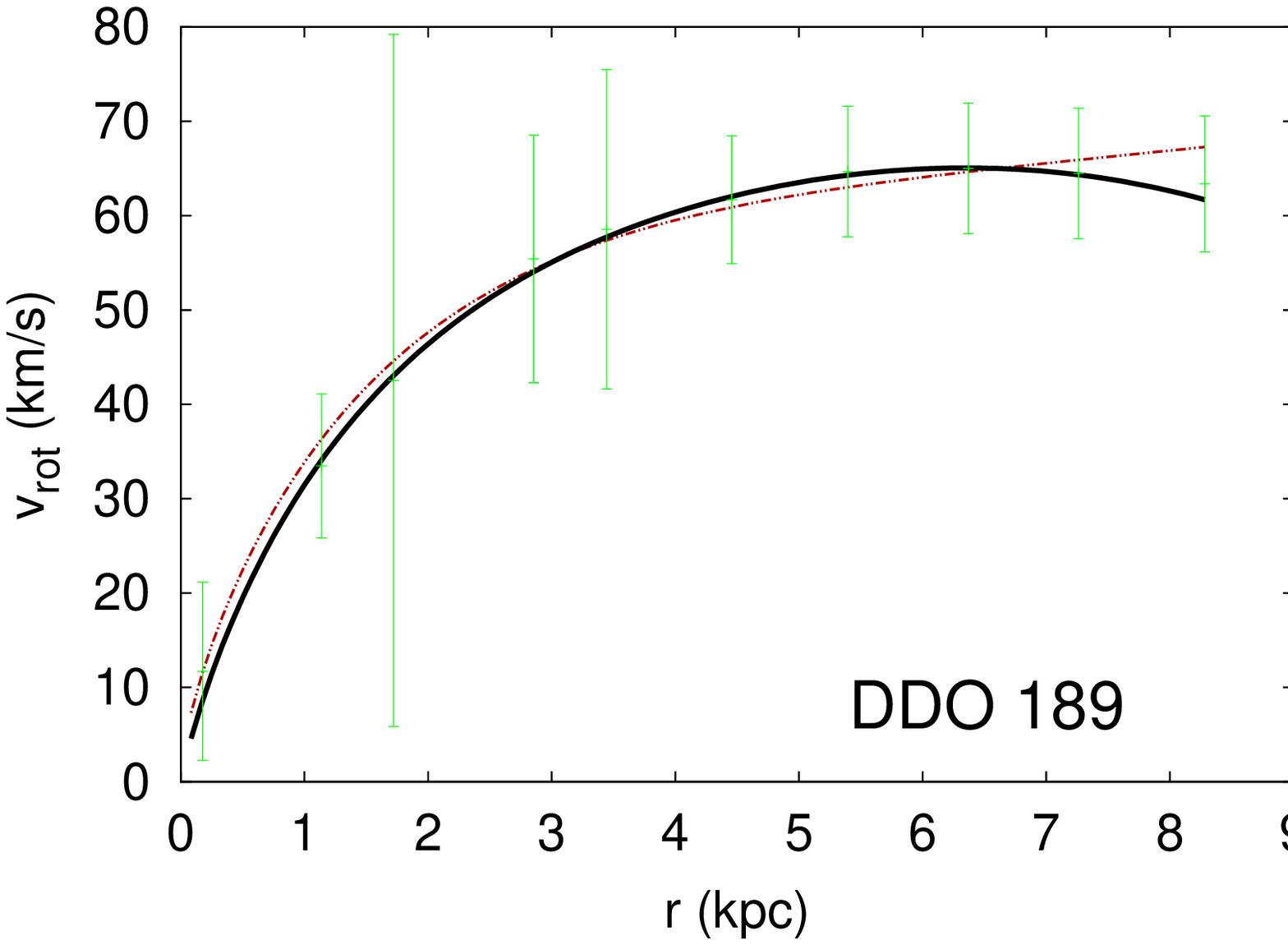} & %
\includegraphics[height=5cm, angle=270]{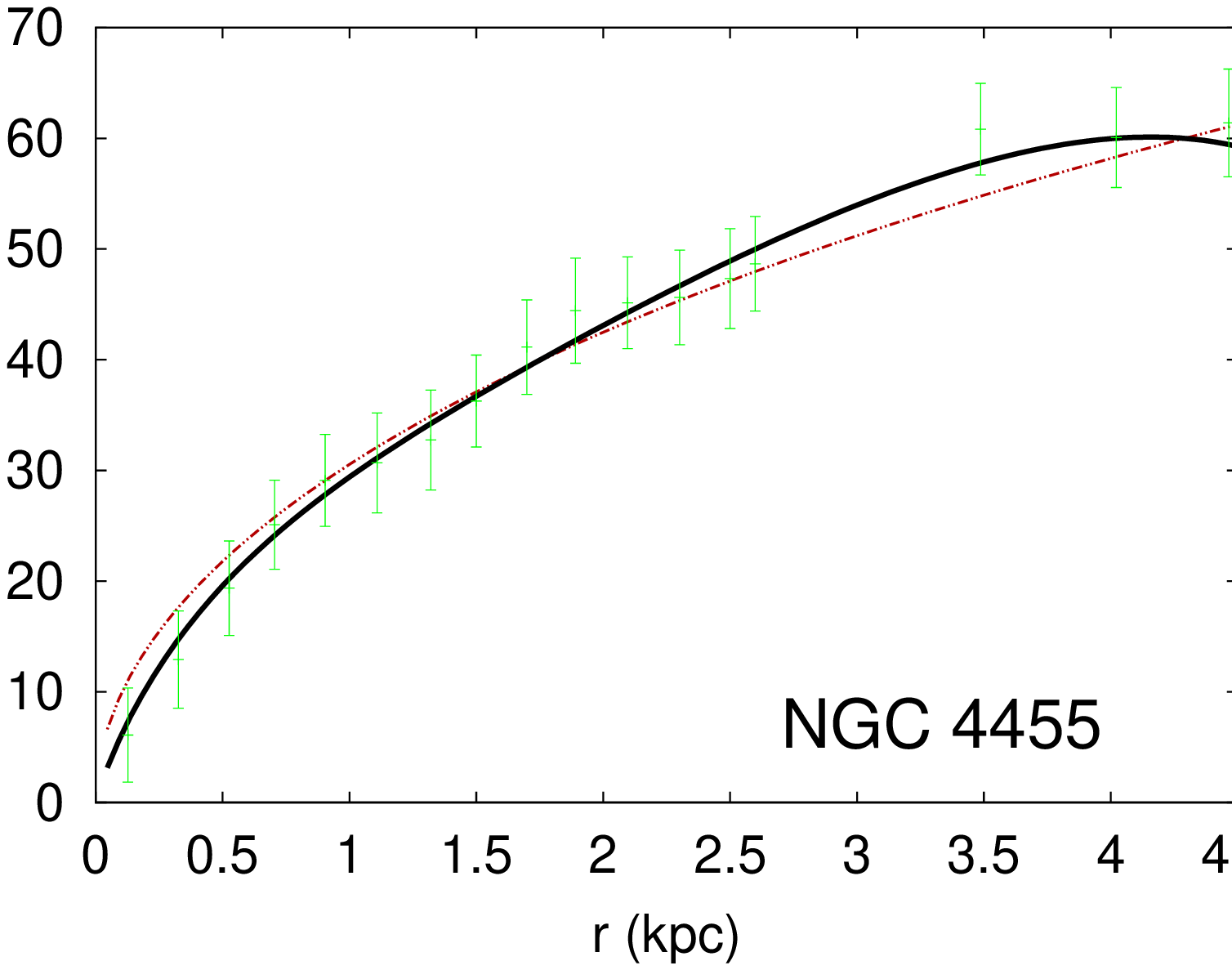} & %
\includegraphics[height=5cm, angle=270]{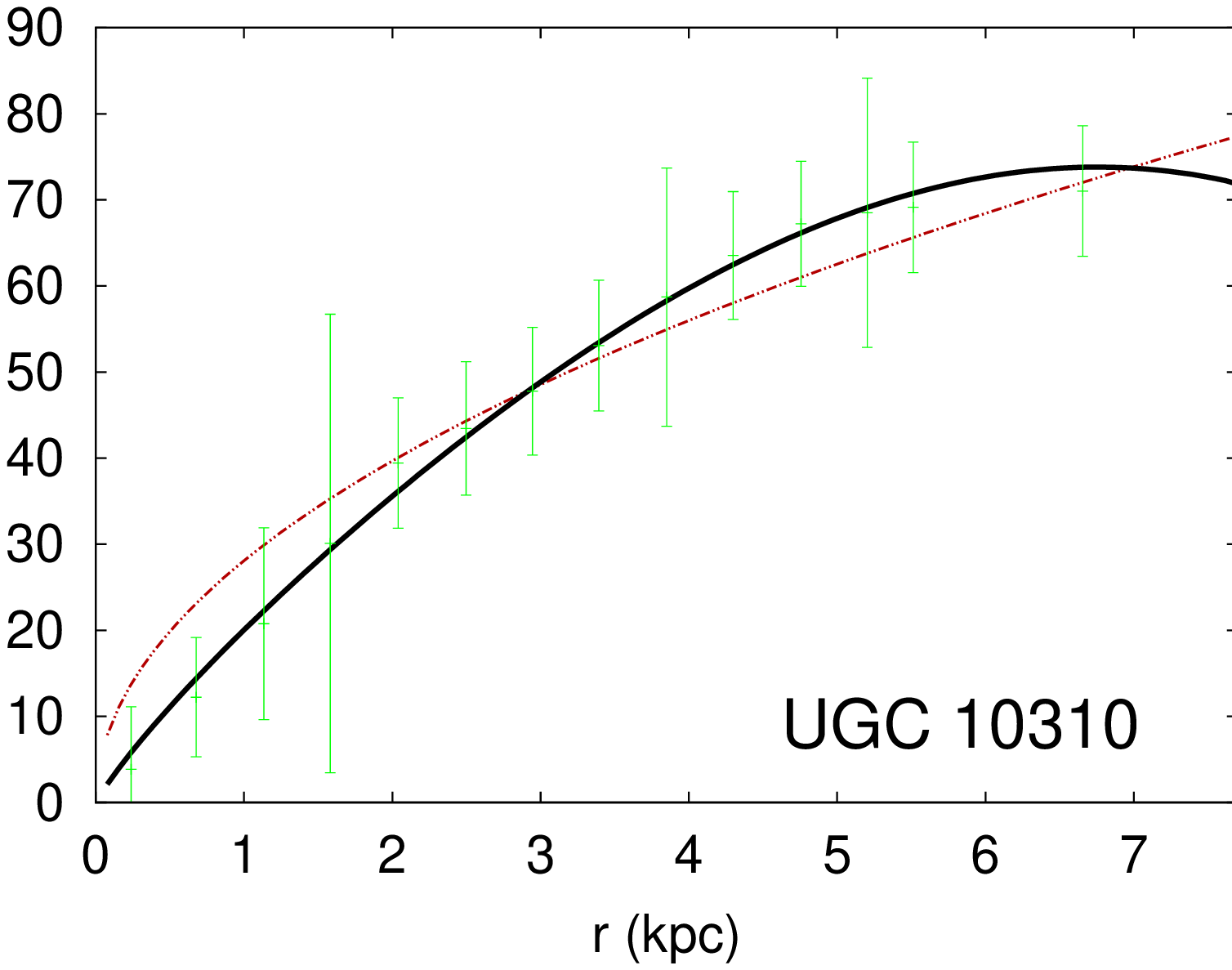} \\ 
&  & 
\end{tabular}%
\end{center}
\caption{Best fit curves for the LSB I. galaxy sample. The solid black lines
indicates the baryonic matter + BEC model, while the dashed red lines indicates the
baryonic matter + NFW model.}
\label{rotcurvLSB1}
\end{figure*}

\begin{figure*}[tbp]
\begin{center}
\begin{tabular}{ccc}
\includegraphics[height=5cm, angle=270]{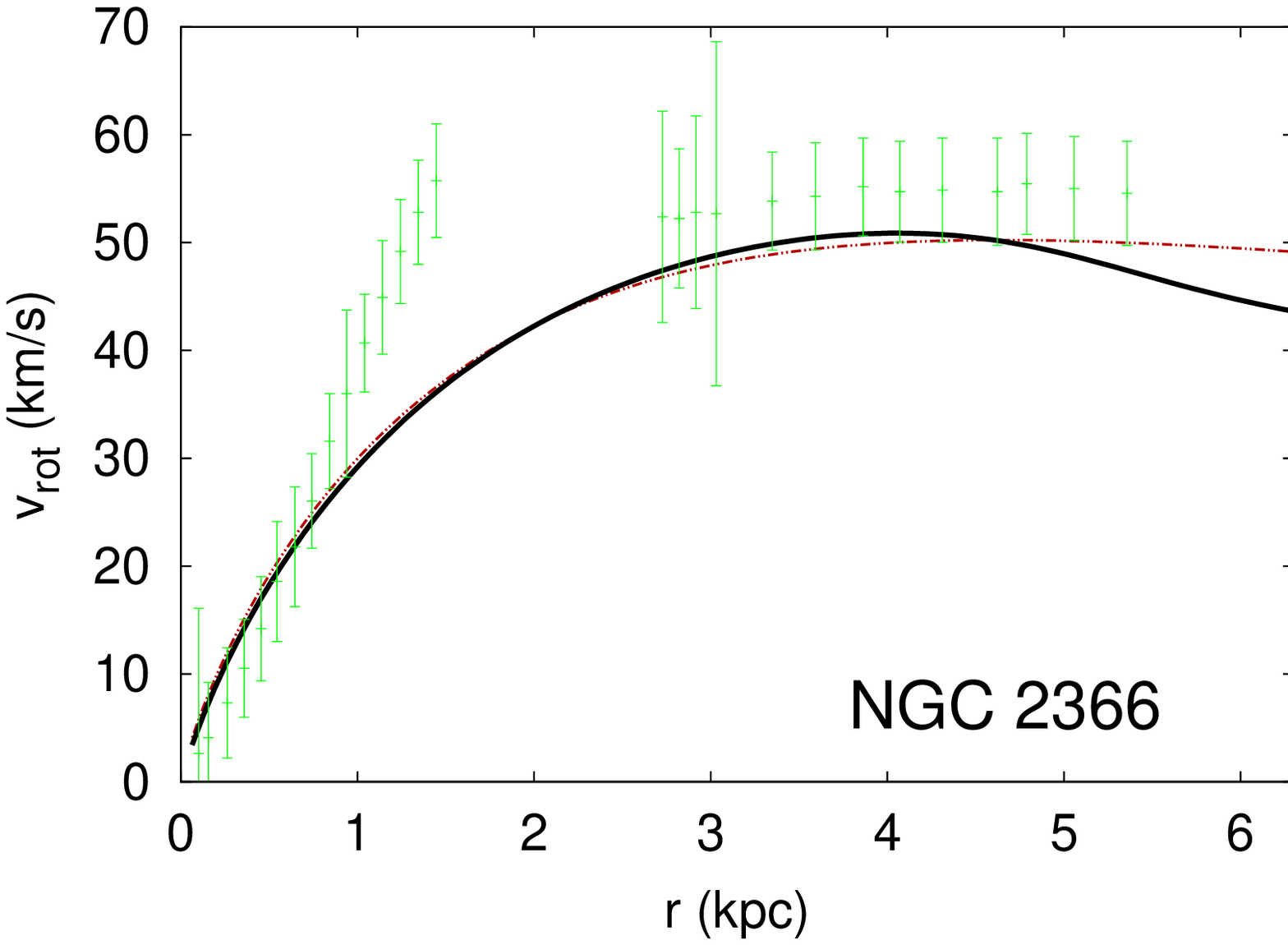} & %
\includegraphics[height=5cm, angle=270]{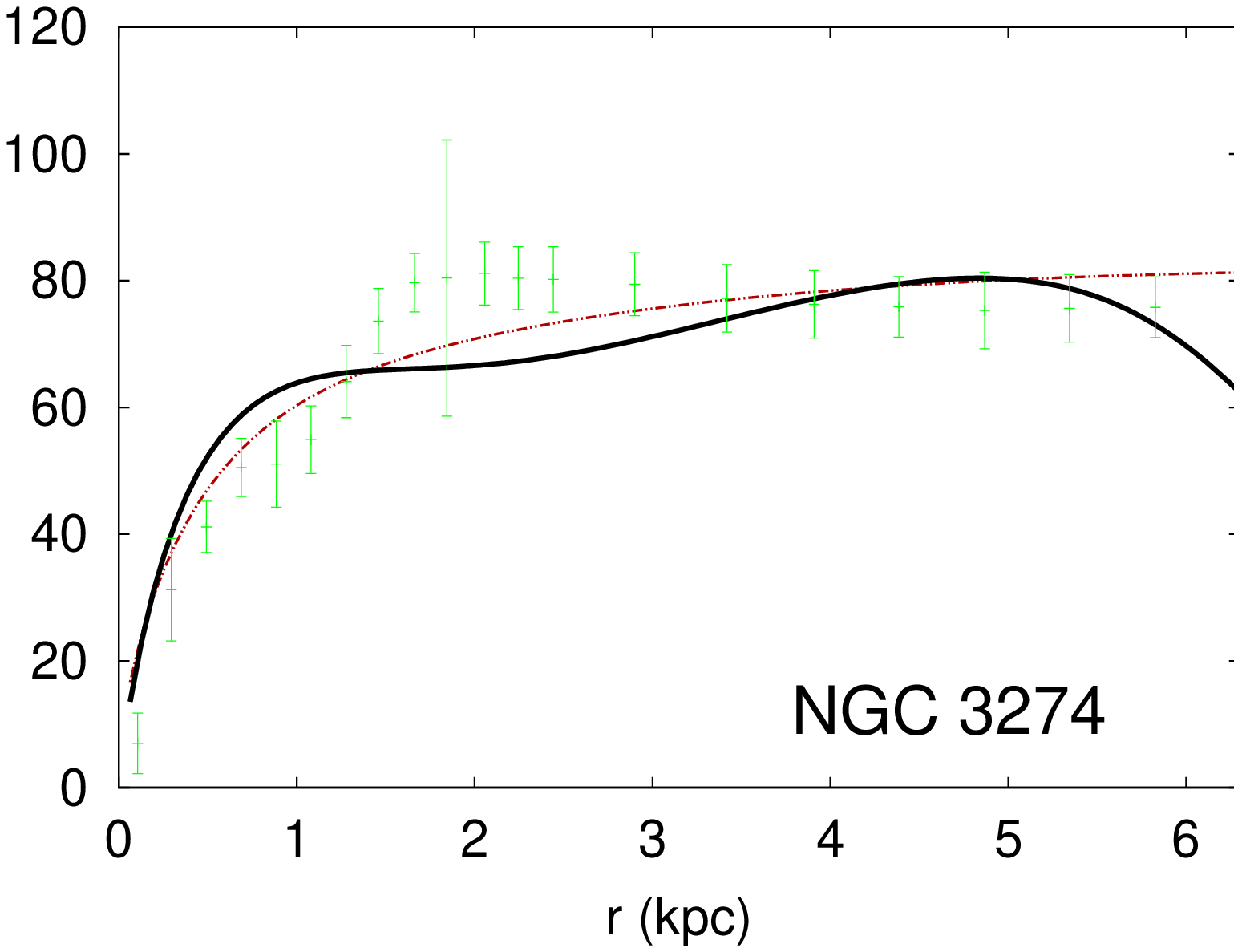} & %
\includegraphics[height=5cm, angle=270]{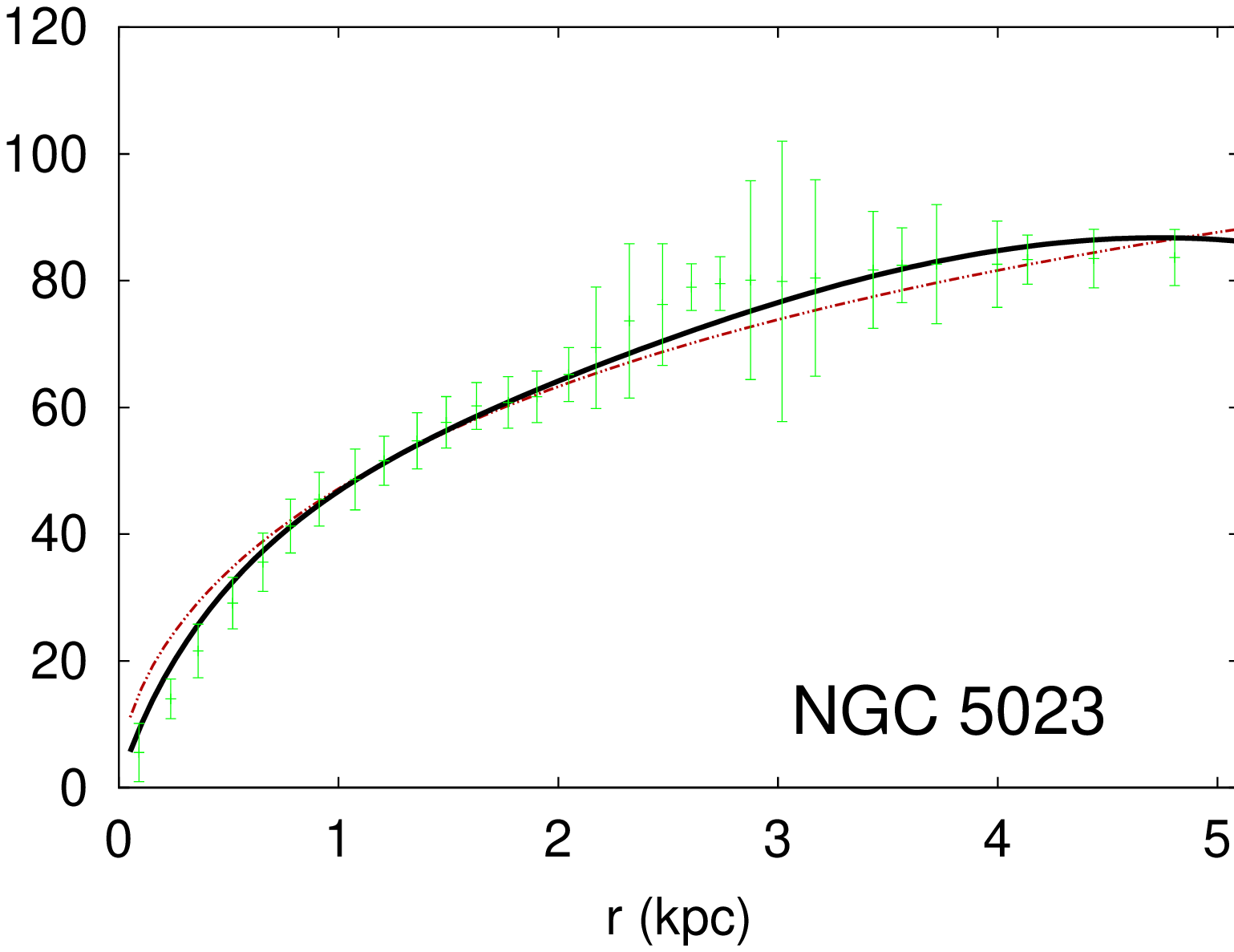} \\ 
&  & 
\end{tabular}%
\end{center}
\caption{Best fit curves for the LSB II. galaxy sample. The solid black
lines refer to the baryonic matter + BEC model, while the dashed red lines
to the baryonic matter + NFW model. As for HSB galaxies, the BEC model fails
to explain the extended flat regions of the rotation curves.}
\label{rotcurvLSB2}
\end{figure*}

We confronted the BEC model with 6 LSB galaxies chosen from a larger sample 
\cite{blok}. The applied data were obtained from  
both $HI$ and $H\alpha$ measurements. From a $\chi ^{2}$-test the
parameters in both the BEC+baryonic and NFW+baryonic models were identified,
these are shown in Table \ref{Tablelsb}. The best fit rotation curves are
represented on Figs.~\ref{rotcurvLSB1} and \ref{rotcurvLSB2}.

\begin{table*}[tbp]
\begin{center}
\resizebox{16.5cm}{!} {
\begin{tabular}{c|c|c|c|c|c|c|c|c|c|c|c}
Galaxy & $D$ & $h^{LSB}$ & $M_{D}^{LSB}(BEC) $ & $\rho^{(c)}_{BEC} $ & $R_{BEC}$ & $\chi _{\min }^{2}(BEC)$
& $M_{D}^{LSB}(NFW) $ & $\rho_{s}$ & $r_{s}$ & $\chi _{\min }^{2}(NFW)$ & 1$\sigma$ \\ \hline
& $Mpc$ & $kpc$ & $10^{9} M_{\odot}$ & $10^{-21}kg/m^3$ & $kpc$ & & $10^{9} M_{\odot}$ & $10^{-24}kg/m^3$ & $kpc$ & & \\ \hline\hline
DDO 189 & 12.6 & 1.9  & 2.71 & 0.38 & 8.3 & 0.519 & 2.16 & 16 & 70 & 1.09 & 7.03 \\
NGC 4455 & 6.8 & 2.3  & 0.231 & 1.44 & 5.5 & 9.29 & 0.11 & 25.04 & 66 & 5.39 & 18.11\\
UGC 10310 & 15.6 & 5.2  & 0.443 & 0.98 & 7.8 & 2.66 & 0.9 & 14.9 & 88 & 5.76 & 13.74\\ \hline\hline
NGC 2366 & 3.4 & 1.5 & 2.43 & 0.22 & 5.3 & 110.73 & 2.5 & 0.2 & 1000 & 116.93 & 26.72 \\
NGC 5023 & 4.8 & 0.8  & 0.894 & 2.45 & 5.6 & 53.2 & 0.0449 & 457 & 13 & 143.08 & 32.05\\
NGC 3274 & 6.7 & 0.5  & 1.1 & 1.69 & 6.4 & 269.8 & 0.252 & 2373 & 4 & 148.44 & 20.27\\

\end{tabular}
}
\end{center}
\caption{The best fit BEC and NFW parameters of the LSB I and II type
galaxies (the first and last three galaxies, respectively). $D$ is taken
from \protect\cite{blok}. The rest of the parameters are rotation curve
fits. For LSB I galaxies the BEC DM model gives significantly better fitting
velocity curves (within 1$\protect\sigma $ confidence level) than the NFW
model. However the velocity curves are outside the 1$\protect\sigma $
confidence level for LSB II galaxies.}
\label{Tablelsb}
\end{table*}

For the LSB I galaxies the BEC DM model gives significantly better fitting
velocity curves (all within the 1$\sigma $ confidence level) compared to the
NFW model (which in two cases out of the three gives fits falling outside 1$%
\sigma $). For LSB II galaxies the
quality of the fits are comparable, but in both models they are beyond the 1$%
\sigma $ confidence level.

\subsection{Dwarf galaxies}

Approximately 85\% of the explored galaxies in the Local Volume \cite{karac} are dwarf
galaxies. The dwarfs are defined by having an absolute magnitude which is fainter than $M_{B}\sim
-16$ $mag$. On the other hand they are larger than globular clusters 
\cite{tamm}.

Although little is known about their formation, it is generally accepted that dwarfs are formed 
at the centres of subhalos. Dwarf galaxies can be categorised in five groups 
according to their optical appearance. The five groups being dwarf
ellipticals, dwarf irregulars, dwarf spheroidals, blue compact dwarfs, and
dwarf spirals. The dwarfs falling in the last group represent the
very small ends of spirals \cite{matt}. Dwarf spheroidals are old systems
and among the most DM dominated galaxies in the Universe.

The central velocity dispersion of most dwarf galaxies is in the range $6\div25$
km/s \cite{mate}. In a typical dwarf galaxy, assuming dynamical equilibrium,
the mass derived from the observed velocity dispersions is substantially greater than
the observed total visible mass. This implies that the mass-to-light ratio
is very high compared to other types of galaxies, hence they can greatly contribute 
to the understanding of DM distribution on small scales. Dwarf
galaxies allow for proving or falsifying different alternative gravity
theories \cite{capo}.

We decided to use 7 dwarf galaxies for testing the BEC model. We have selected 
the sample dwarf galaxies such as to ensure that sufficient high resolution rotation 
curve data would be available for our study. 
We fitted both the BEC+baryonic and the NFW+baryonic models, respectively, with
similar baryonic components as for the LSB galaxies. As the length scales of
the stellar disks were not available for the selected sample, they were calculated by 
$\chi ^{2}$ minimization, too.

\begin{figure*}[tbp]
\begin{center}
\begin{tabular}{ccc}
\includegraphics[height=5cm, angle=270]{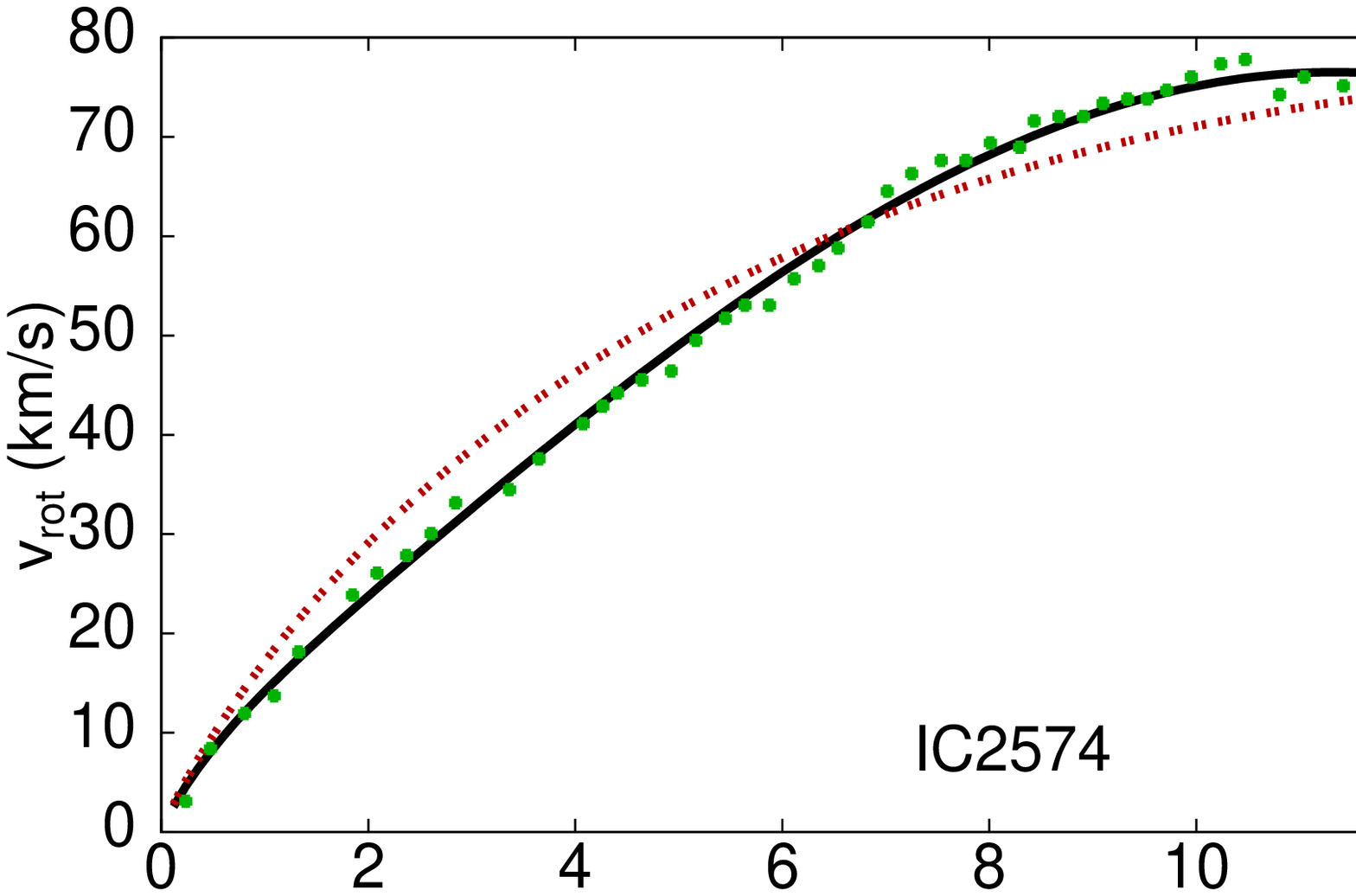} & %
\includegraphics[height=5cm, angle=270]{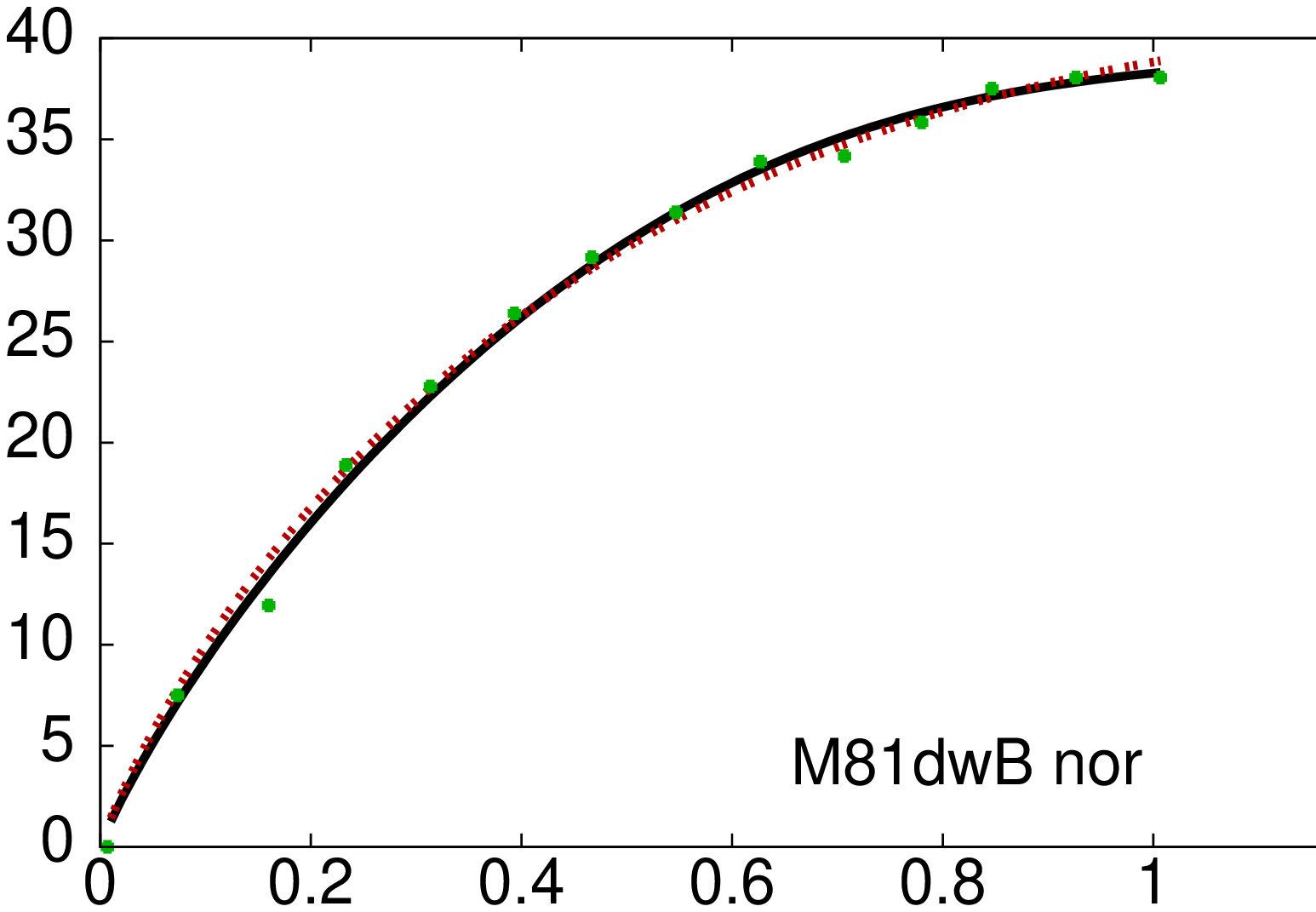} & %
\includegraphics[height=5cm, angle=270]{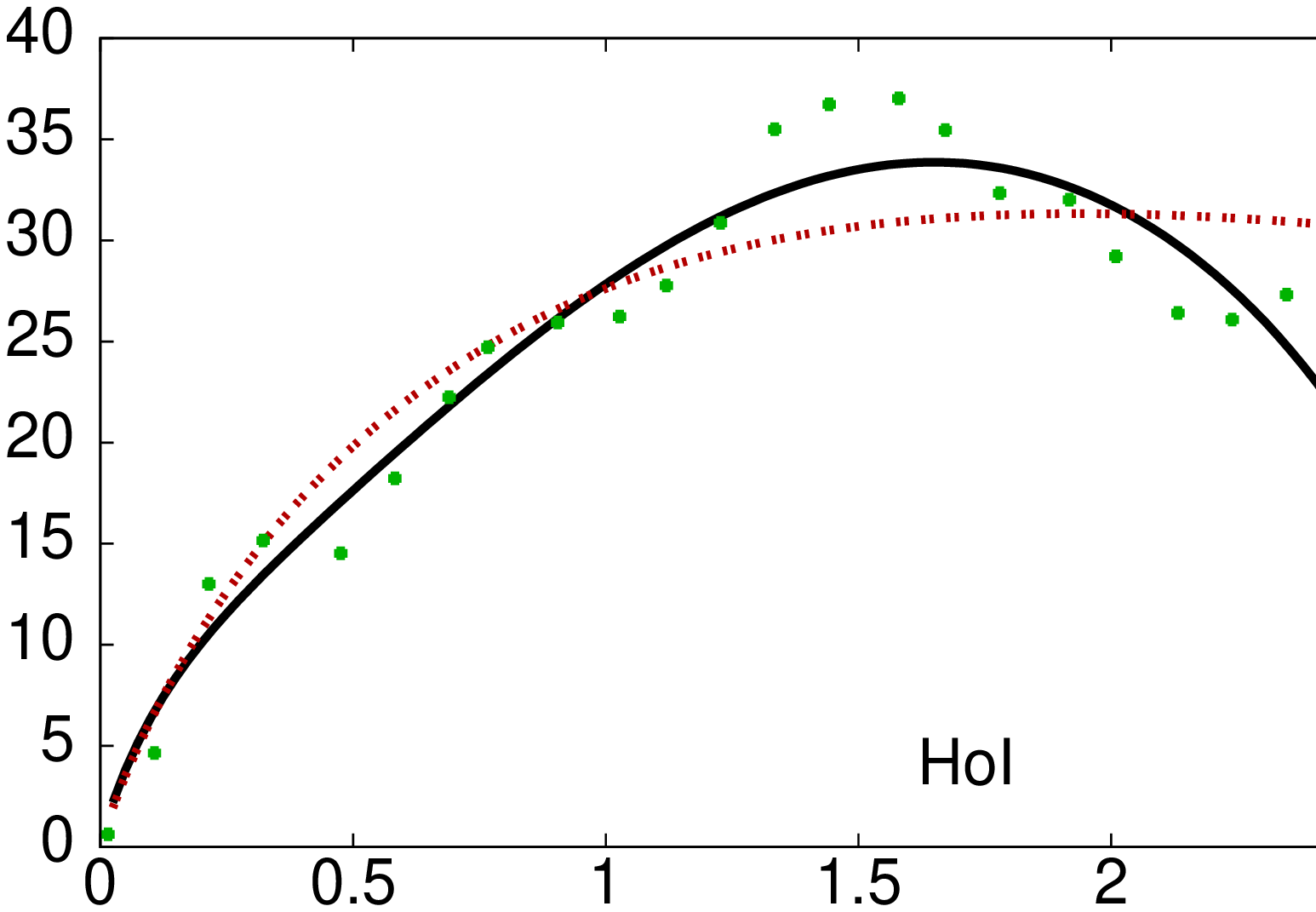} \\ 
\includegraphics[height=5cm, angle=270]{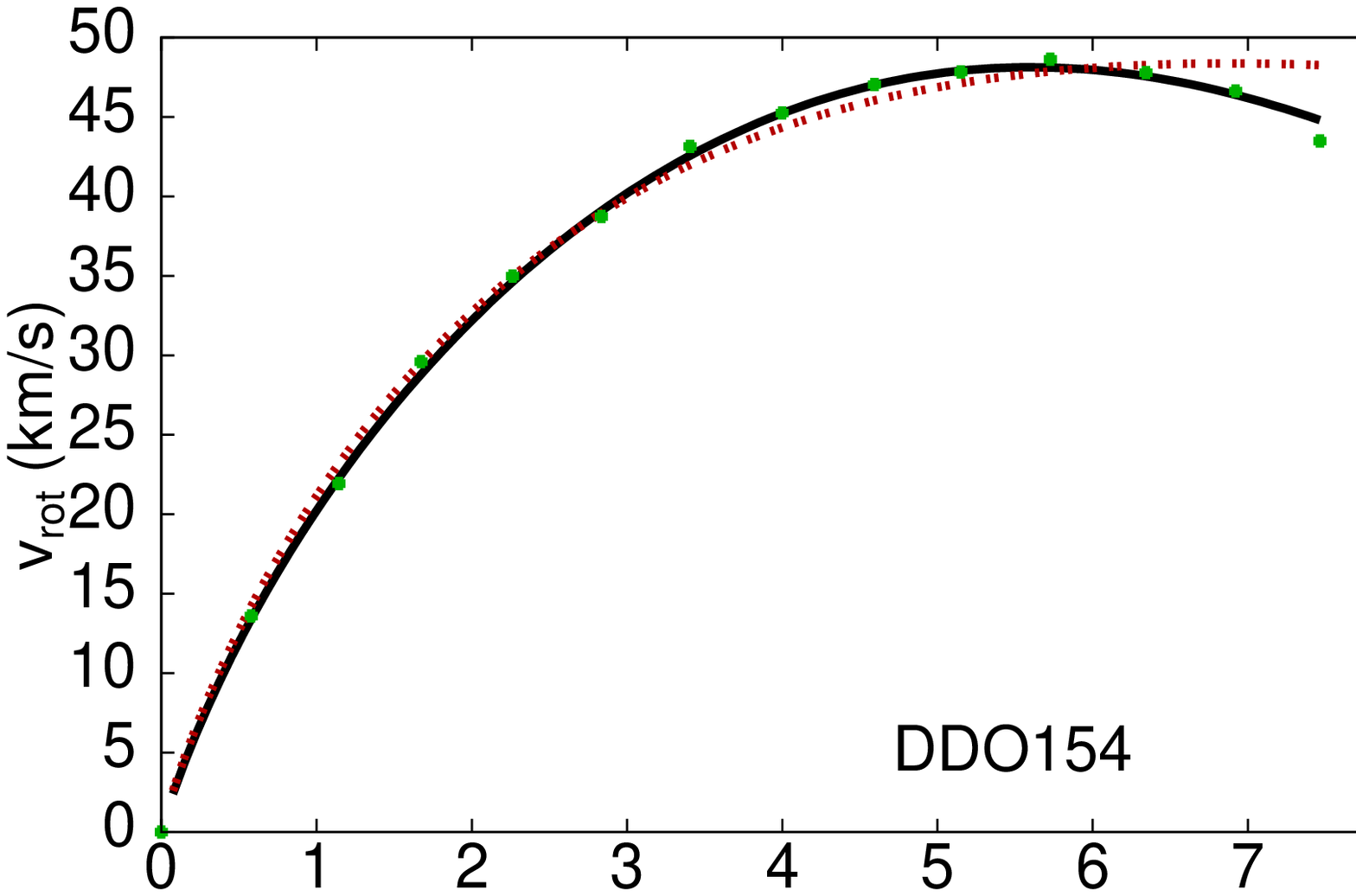} & %
\includegraphics[height=5cm, angle=270]{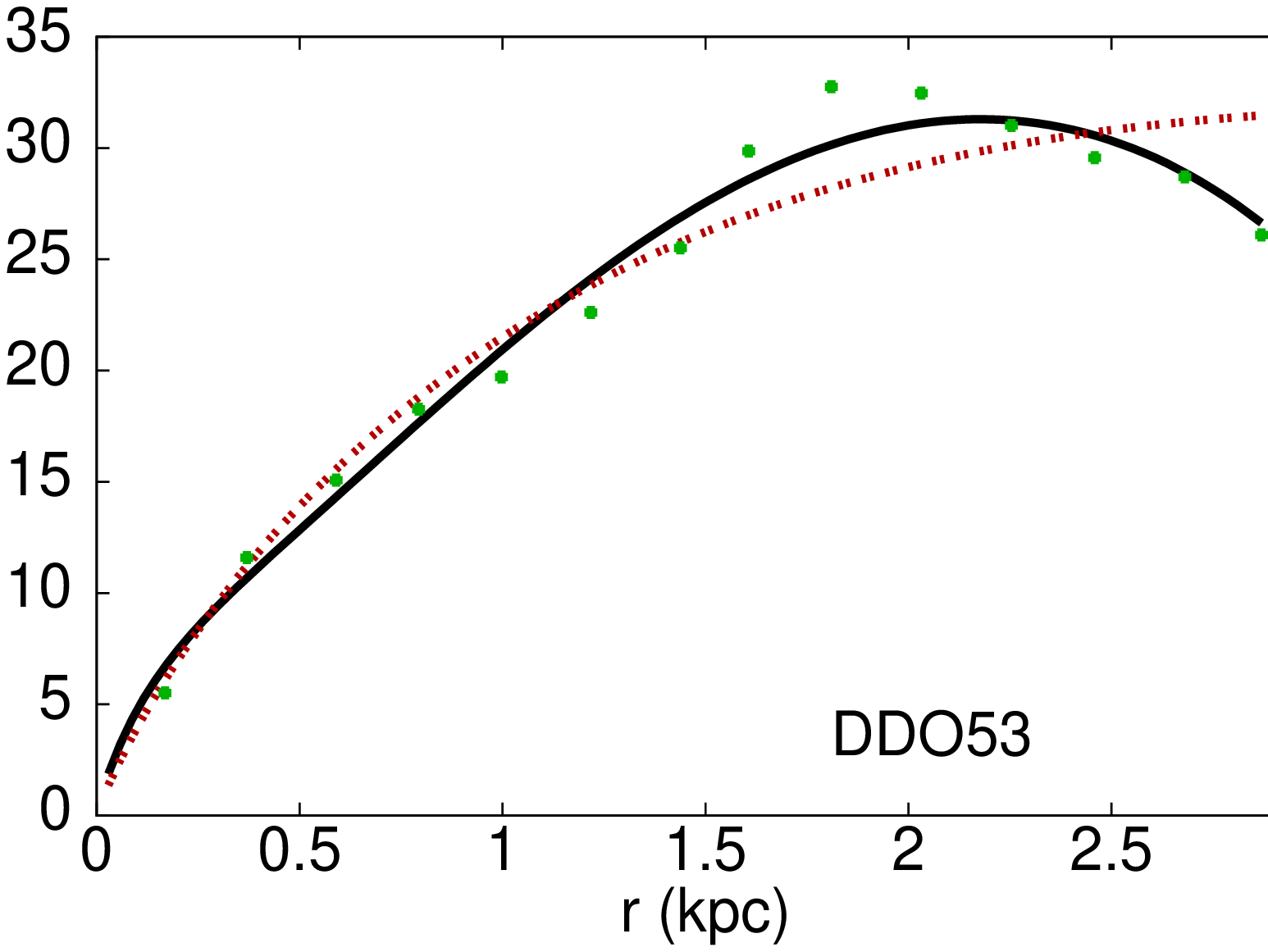} & %
\includegraphics[height=5cm, angle=270]{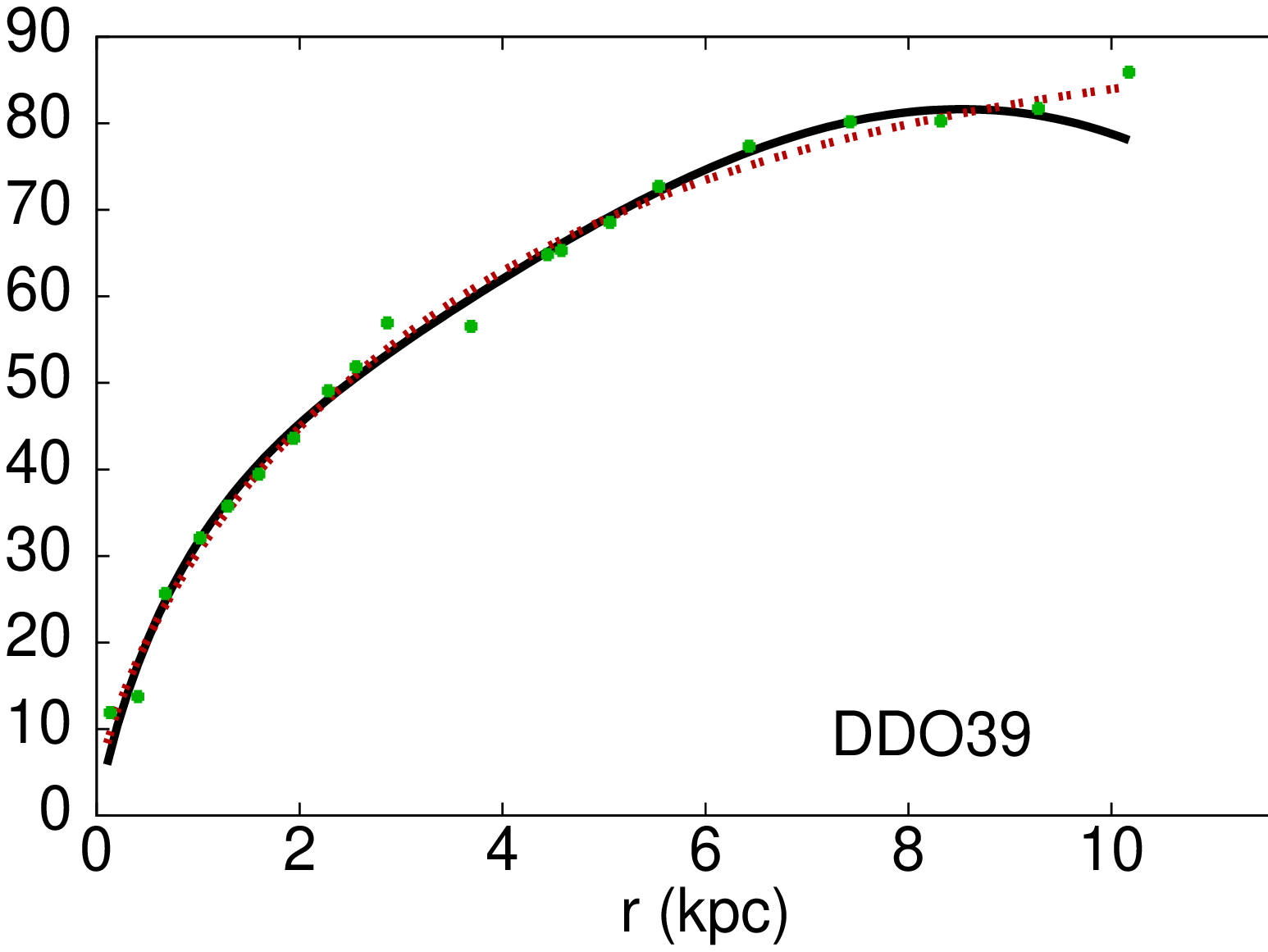} \\ 
\includegraphics[height=5cm, angle=270]{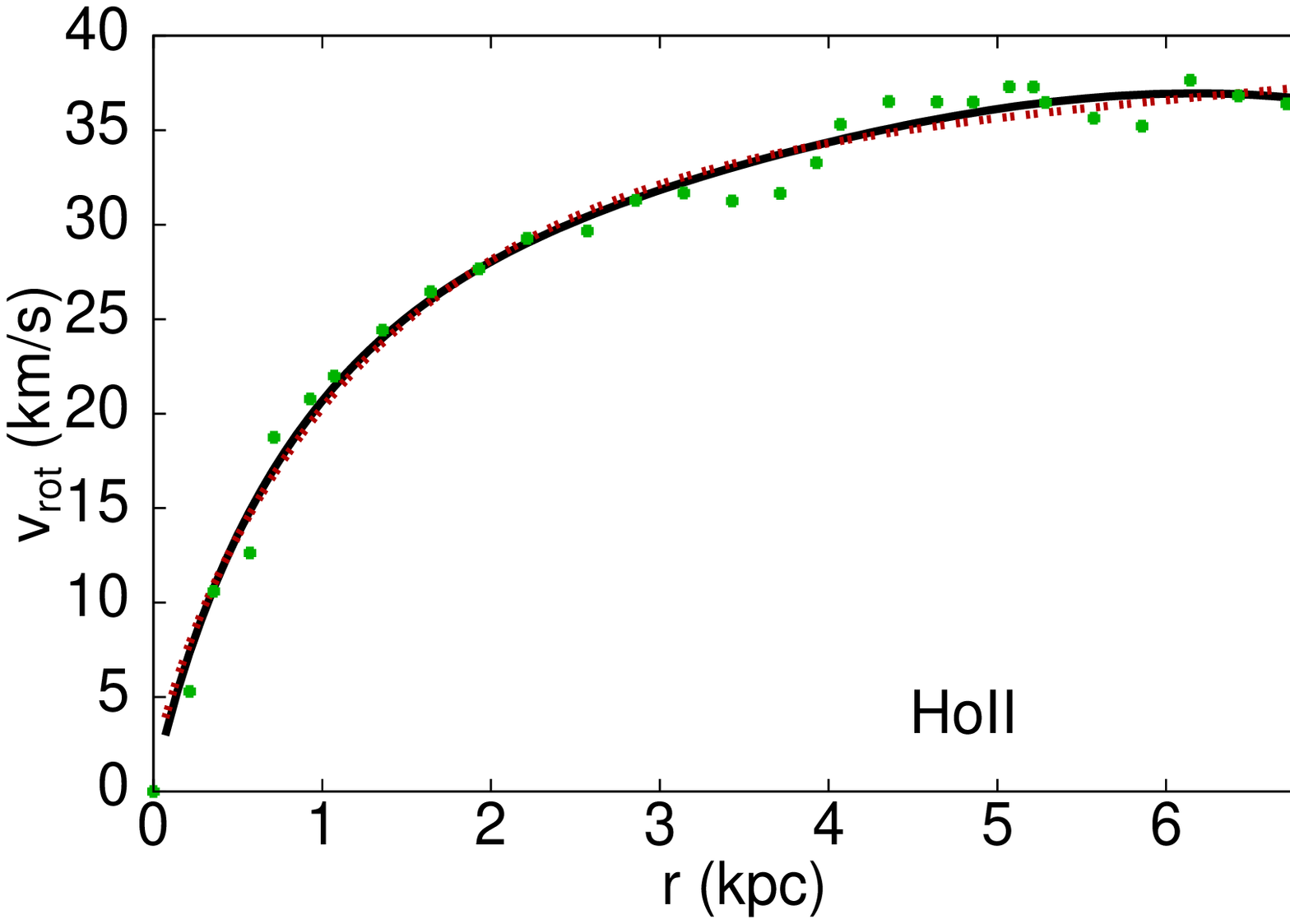} &  & 
\end{tabular}%
\end{center}
\caption{The best fit curves for the dwarf galaxy sample. The BEC+baryonic
model (solid black curves) gives a better fit in all cases then the
NFW+baryonic model (dashed red lines). In both cases the fit was performed
with the same baryonic model.}
\label{dwarfcombined}
\end{figure*}

A preliminary check showed that the addition of the BEC dark matter halo to
the baryonic model improved (giving lower $\chi _{\min }^{2}$ values) on the
fit in all cases. By contrast, the NFW\ model was unable to improve on the
purely baryonic fit in four out of seven cases. We note that since the data
does not contain the error margins, the $\chi _{\min }^{2}$ values are
relatively high (beyond the 1$\sigma $ confidence level in most cases). The
best fit BEC and NFW parameters are shown in Table~\ref{Tabledwarfbec} and
the corresponding rotation curves are represented on Fig.~\ref{dwarfcombined}%
. The inclusion of the BEC DM model gives significantly (in some cases one
order of magnitude in the value of $\chi^{2}$) better fits compared to the
case of NFW model. This is due to the cusp avoidance in the central density
profile of the BEC model and the fact that dwarf galaxies do not exhibit
extended flat regions in their rotation curves.

\begin{table*}[tbp]
\begin{center}
\resizebox{16.5cm}{!} {
\begin{tabular}{c|c|c|c|c|c|c|c|c|c|c|c}
Galaxy & $h^{dwarf}(BEC)$ & $M_{D}^{dwarf}(BEC) $ & $\rho^{(c)}_{BEC} $ & $R_{BEC}$ & $\chi _{\min }^{2}(BEC)$
& $h^{dwarf}(NFW)$ & $M_{D}^{dwarf}(NFW) $ & $\rho_{s}$ & $r_{s}$ & $\chi _{\min }^{2}(NFW)$ & 1$\sigma$ \\ \hline
& $kpc$ & $10^{9} M_{\odot}$ & $10^{-21}kg/m^3$ & $kpc$ & & $kpc$& $10^{9} M_{\odot}$ & $10^{-24}kg/m^3$ & $kpc$ & & \\ \hline\hline
IC 2574 & 1.2 & 0.1122 & 0.4 & 13 & 68.47 & 7.9 & 28.44 & 0 & 0 & 714.73 & 44.74  \\
HoI & 0.2 & 0.0107  & 3.6 & 1.9 & 95.26 & 0.9 & 0.533 & 0 & 0 & 241.30 & 20.27 \\
HoII & 1.2 & 0.4431  & 0.2 & 7.69 & 33.33 & 1.7 & 0.642 & 4 & 92 & 43.86 & 26.72 \\
DDO 39 & 1.3 & 1.1235  & 0.7 & 10.01 & 69.39 & 4.3 & 7.21 & 43 & 35 & 69.82 & 17.02 \\
DDO 53 & 0.2 & 0.0061  & 1.8 & 2.5& 20.05 & 1.6 & 0.976 & 1 & 24 & 51.53 & 10.42 \\
DDO 154 & 3.1 & 3.3502  & 0.2 & 5.8 & 1.48 & 3.2 & 4.52 & 0 & 0 & 9.40 & 9.30 \\
M81dwB nor & 0.9 & 1.023  & 3.7 & 0.7 & 6.19 & 0.7 & 0.705 & 0 & 0 & 8.4 & 10.42 \\
\end{tabular}
}
\end{center}
\caption{The best rotation curve data fit BEC+baryonic and NFW+baryonic
parameters for the dwarf galaxy sample.}
\label{Tabledwarfbec}
\end{table*}

\section{Discussions and final remarks}
\label{section4}

We have performed a $\chi ^{2}$-test of the BEC and NFW DM models, with the
rotation curves of 6 HSB, 6 LSB and 7 dwarf galaxy samples. For
improved accuracy we also included realistic baryonic models in every case.
For the HSB galaxy sample, both the rotation curve and the surface
photometry data were available. Most of the rotation curves were smooth,
symmetric and uniform in quality.

The circular velocity of the investigated galaxies was decomposed into 
its barionic and DM contribution: 
$v_{model}^{2}(r)=v_{baryonic}^{2}+v_{DM}^{2}$. For the BEC model the DM
contribution to the rotational velocity can be described as Eq.~(\ref{vel}). Then the rotation
curves are fitted with the parameters of the baryonic and DM halo models (BEC and NFW) using 
$\chi^{2}$ minimization method.

The analysis of the \textit{HSB I galaxies} showed a remarkably good
agreement for both DM models with observations. The BEC and NFW models show similar fits. 
However, the rotation curves of the HSB II type galaxies are significantly better
described by the NFW model.

It was previously known that for \textit{LSB galaxies and without including
the baryonic sector}, the BEC model gave a better fit than the NFW model 
\cite{robl}. We additionally found that including the baryonic component
improves on the fit of \cite{robl}. Our detailed analysis showed a significantly better performance of the BEC model for LSB type I galaxies, while comparable fits for LSB type II galaxies were obtained. These latter fits were
however outside the 2$\sigma$ confidence level.

The unsatisfactory large distance behaviour of the BEC model for both the
HSB and LSB galaxies of type II originates in the sharp cutoff of the BEC DM
distribution and clearly indicates that it would be desirable to modify the BEC model on
larger scale, also to comply with the 
behaviour of the universal rotation curves (URCs) at larger radii \cite{pers}.

From the above analysis of HSB and LSB galaxies it is also obvious that
(while on large distances the BEC model suffers from problems due to the
sharp cutoff) close to the core it works overall better than the NFW model. This
is also supported by our fit of both the BEC+baryonic and NFW+baryonic DM
models with rotation curve data of a sample of 7 dwarf galaxies. Since dwarf
galaxies are DM dominated, they allow for the best comparison between the various models. 
The results can be seen in Fig. \ref{dwarfcombined}. We also note that
the NFW DM improved over the pure baryonic fit in four cases out of seven,
while including the BEC component improved over the fit with the baryonic
component in all cases.

The BEC parameters were determined for all cases. The parameters $\rho _{BEC}^{(c)}$, $%
R_{BEC}$ are given in Tables \ref{Tablehsbpar}, \ref{Tablelsb}, \ref%
{Tabledwarfbec}. The averages of the radii $R_{DM}$ of the
BEC halos for the HSB, LSB and dwarf galaxies are $\langle
R_{BEC}^{HSB}\rangle \approx 4.06\mathrm{kpc}$, $\langle R_{DM}^{LSB}\rangle
\approx 6.48\mathrm{kpc}$ and $\langle R_{DM}^{dwarf}\rangle \approx 5.94kpc$%
, respectively. The scatter however is large, there are no universal BEC
parameters which globally fit all the galaxies, not even at 3$\sigma $
confidence level. The closer to this goal were the HSB galaxies, where 3 out
of 6 had overlapping 3$\sigma $ domains. Nonetheless the given values of $%
R_{DM}$ are consistent within the order of magnitude with the halo radii of
59 other galaxies determined from weak lensing \cite{pire}.

\begin{figure*}[tbp]
\begin{center}
\begin{tabular}{cc}
\includegraphics[height=5cm]{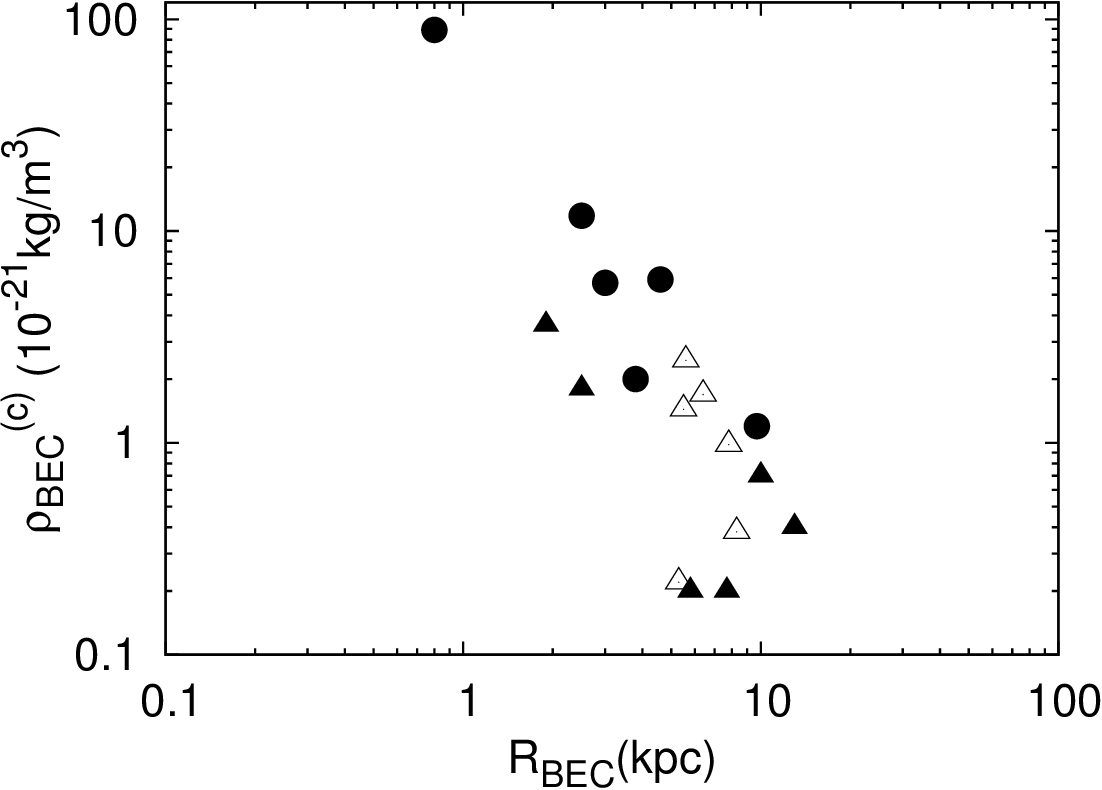} & %
\includegraphics[height=5cm]{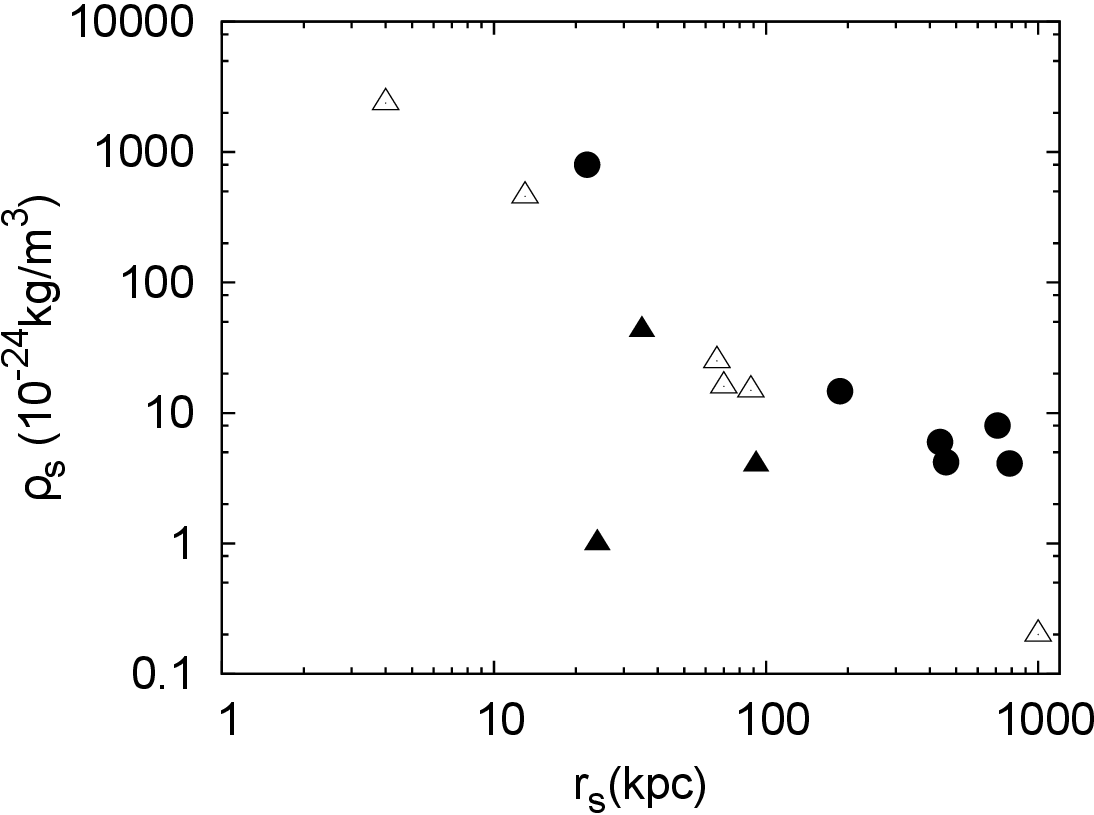} \\ %
\end{tabular}%
\end{center}
\caption{The density parameter $\rho^{(c)}_{BEC} $ of the BEC model is shown as function of $R_{BEC}$ in the left panel, and the density parameter $\rho_{s} $ of the NFW model is shown as function of $r_{s}$ in the right panel. The HSB, LSB, dwarf galaxies are represented by filled circles, empty triangles, and filled triangles, respectively.}
\label{rhor} 
\end{figure*}

We represent the density parameter $\rho^{(c)}_{BEC} $ of the BEC model as function of $R_{BEC}$ in the left panel of Fig. \ref{rhor}, and the density parameter $\rho_{s}$ of the NFW model as function of $r_{s}$ in the right panel of Fig. \ref{rhor} (four dwarf galaxies are absent, as the NFW halo does not improve the fit over the pure baryonic case). The fitting enforces a relation between the dark matter parameters: the characteristic density scales inversely with the corresponding characteristic distance.

\begin{figure*}[tbp]
\centering
\includegraphics[height=5cm]{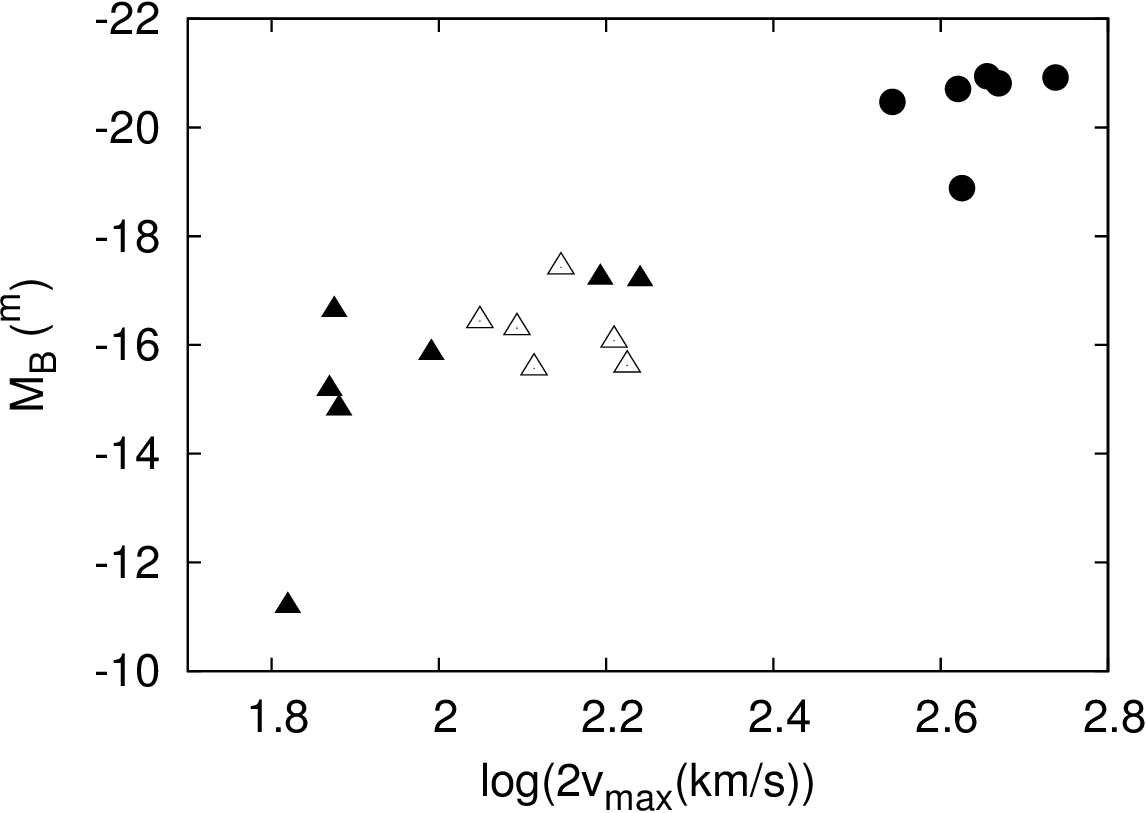}
\caption{The baryonic Tully-Fisher relation of our galaxy sample. Absolute B magnitudes are presented as function of the logarithm of the maximal rotational velocity. The HSB, LSB, dwarf galaxies are represented by filled circles, empty triangles, and filled triangles, respectively.}
\label{tullyfisher} 
\end{figure*}

We verify the Tully-Fisher relation for the investigated galaxy sample, and present the results on Fig. \ref{tullyfisher}. Apparent B magnitudes and galaxy distances were collected from the NASA/IPAC extragalactic database \cite{Helou1991}, and were corrected for extinction based on Landolt standard-fields to calculate the absolute magnitudes. It is known that the Tully-Fisher relation holds for spiral and lenticular galaxies with the same slope (e.g \cite{McGaugh2000}). A larger slope and scatter characterize the Tully-Fisher relation for the dwarf galaxies (e.g \cite{McGaugh2000,McGaugh2010}). The investigated sample exactly exhibits these features.

There is a relation among the mass $m$ of the BEC particle, its coherent
scattering length $a$ and the radius of the DM halo $R_{DM}$ \cite{boeh1}: 
\begin{eqnarray}
m &=&\left( \frac{\pi ^{2}\hbar ^{2}a}{GR_{BEC}^{2}}\right) ^{1/3}  \nonumber
\\
&\approx &6.73\times 10^{-2}\left[ a(\mathrm{fm})\right] ^{1/3}\left[
R_{BEC}\;\mathrm{(kpc)}\right] ^{-2/3}\mathrm{eV}.
\end{eqnarray}%
Axions have been proposed as the Peccei-Quinn solution to the strong CP problem \cite{PecceiQuinn} and they are among the best dark matter candidates. Being bosons, they may also form BEC. The Axion Dark Matter experiment has already established limits on the dark matter axions \cite{axionL1,axionL2}.

Assuming the BEC is formed of axions with mass of $10^{-6}\mathrm{eV}$, the
scattering lengths for the three types of galaxies emerge as $a_{HSB}\approx
5.4\times 10^{-14}\mathrm{fm}$, $a_{LSB}\approx 1.37\times 10^{-13}\mathrm{fm%
}$ and $a_{dwarf}\approx 1.15\times 10^{-13}fm$. These values are consistent
with the results of \cite{pire}, which are based on a statistical analysis
of 61 DM dominated galaxies. The total energy of the BEC halo is negative
with these scattering lengths and particle mass, meaning the halo is stable
(see Fig. 3 of \cite{souza14}).

\ack In the earlier stages of this work MD, ZK and L\'{A}G were
supported by the European Union and the State of Hungary, co-financed by the
European Social Fund in the framework of T\'{A}MOP 4.2.4.
A/2-11-/1-2012-0001 `National Excellence Program'. L\'{A}G was also
supported by the Japan Society for the Promotion of Science.


\section*{References}

\begin{thebibliography}{99}

\bibitem{planck1} Ade P A R, Aghanim N, Armitage-Caplan C, et al., \textit{%
Planck 2013 results. I. Overview of products and scientific results}, 2014 \textit{A\&A} \textbf{571}, 1

\bibitem{planck2} Francis M, \textit{First Planck results: the Universe is
still weird and interesting}, 2013 \textit{Arstechnica}

\bibitem{jaloc} Ja\l ocha J, Bratek \L , Kutschera M and Skindzier P., 
\textit{Global disk model for galaxies NGC 1365, NGC 6946, NGC 7793, UGC 6446%
}, 2010 \textit{MNRAS} \textbf{406}, 2805-2816

\bibitem{milg} Milgrom M, \textit{A Modification of the Newtonian dynamics
as a possible alternative to the hidden mass hypothesis}, 1983 \textit{ApJ} 
\textbf{270}, 365

\bibitem{sand} Sanders R H, \textit{Anti-gravity and galaxy rotation curves}%
, 1984 \textit{A\&A} \textbf{136}, L21

\bibitem{mof} Moffat J W and Sokolov I Y, \textit{Galaxy dynamics
predictions in the nonsymmetric gravitational theory}, 1996 \textit{Phys.
Lett. B} \textbf{378}, 59

\bibitem{mann} Mannheim P D, \textit{Are galactic rotation curves really
flat?}, 1997 \textit{ApJ} \textbf{479}, 659

\bibitem{roberts} Roberts M D, \textit{Galactic metrics}, 2004 \textit{Gen.
Rel. Grav.} \textbf{36}, 2423

\bibitem{boeh3} Boehmer C G and Harko T, \textit{On Einstein clusters as
galactic dark matter halos}, 2007 \textit{MNRAS} \textbf{379}

\bibitem{boeh1} Boehmer C G, Harko T, \textit{Can dark matter be a
Bose-Einstein condensate?}, 2007 \textit{JCAP} \textbf{06}, 025

\bibitem{berto} Bertolami O, Boehmer C G, Harko T and Lobo F S N, \textit{%
Extra force in f(R) modified theories of gravity }, 2007 \textit{Phys. Rev. D} 
\textbf{75}, 104016

\bibitem{boeh2} Boehmer C G, Harko T, Lobo F S N, \textit{Dark matter as a
geometric effect in f(R) gravity}, 2008 \textit{Astropart. Phys.} \textbf{29}%
, 386

\bibitem{mak} Mak M K and Harko T, \textit{Can the galactic rotation curves
be explained in brane world models?}, 2004 \textit{Phys. Rev. D} \textbf{70}%
, 024010

\bibitem{raha} Rahaman F, Kalam M, DeBenedictis A, Usmani A A and Saibal R, 
\textit{Galactic rotation curves and brane world models}, 2008 \textit{MNRAS}
\textbf{389}, 27

\bibitem{gerg} Gergely L \'{A}, Harko T, Dwornik M, Kupi G and Keresztes Z, 
\textit{Galactic rotation curves in brane world models} 2011 \textit{MNRAS} 
\textbf{415}, 3275

\bibitem{stabile} Stabile A and Capozziello P,  2013 \textit{Phys. Rev. D} \textbf{87}, 064002

\bibitem{supersymmetricLHC1} Conover E, \textit{Supersymmetry’s absence at LHC puzzles physicists}, 2016, textit{ScienceNews} \textbf{190} (7) 12

\bibitem{supersymmetricLHC2} N. Arkani-Hamed et al. \textit{SUSY Bet: Arkani-Hamed and Panel Discussion}, Current Themes in High Energy Physics and Cosmology, Copenhagen, Denmark, 2016

\bibitem{SterileNeutrinoIceCube} Aartsen M G et al. (IceCube Collaboration), \textit{Searches for Sterile Neutrinos with the IceCube Detector}, 2016, textit{Phys. Rev. Lett.} \textbf{117}, 071801

\bibitem{Lux} LUX Collaboration, \textit{First results from the LUX dark matter experiment at the Sanford Underground Research Facility}, 2013 \textit{arXiv:1310.8214}

\bibitem{Panda2016} PandaX-II Collaboration, \textit{Dark Matter Results from First 98.7-day Data of PandaX-II Experiment}, 2016 \textit{arXiv:1607.07400}

\bibitem{Xenon100} XENON100 Collaboration, \textit{XENON100 Dark Matter Results from a Combination of 477 Live Days}, 2016 \textit{arXiv:1609.06154}

\bibitem{Choudhury} Choudhury D and {Ghosh} K, \textit{Bounds on universal extra dimension from LHC run I and II data}, 2016 \textit{Phys. Rev. Lett. B} \textbf{763}, 155-160

\bibitem{Macho} {Alcock} C et al., \textit{The MACHO Project: Microlensing Results from 5.7 Years of Large Magellanic Cloud Observations}, 2000 \textit{ApJ} \textbf{542}, 281-307

\bibitem{GW} {Abbott} B. P. et al. \textit{Observation of Gravitational Waves from a Binary Black Hole Merger}, 2016 \textit{Phys. Rev. Lett.} \textbf{116}, 6

\bibitem{primack} Primack J R and Gross M A K, \textit{Current Aspects of
Neutrino Physics (Springer, Berlin Heidelberg 2000)}, 2000

\bibitem{vega} de Vega H J and Sanchez N G, \textit{Warm dark matter in the
galaxies:theoretical and observational progresses. Highlights and
conclusions of the chalonge meudon workshop 2011}, 2011 \textit{%
arXiv:1109.3187}

\bibitem{weietal} Wei H, Chen Z and Liu J, \textit{Cosmological Constraints
on Variable Warm Dark Matter}, 2013 \textit{Phys. Lett. B} \textbf{720},
271-276

\bibitem{biermann} Biermann P L and Kusenko A, \textit{Relic keV sterile
neutrinos and reionization}, 2006 \textit{Phys.Rev.Lett.} \textbf{96}, 091301

\bibitem{IceCUBE} IceCube Collaboration: Aartsen M G et al., \textit{%
Searches for Sterile Neutrinos with the IceCube Detector}, 2016 \textit{%
Phys.Rev.Lett.} \textbf{117}, 071801

\bibitem{padm} Padmanabhan T, \textit{Cosmological constant: The Weight of
the vacuum}, 2003 \textit{Phys. Repts.} \textbf{380}, 235

\bibitem{peeb} Peebles P J E and Ratra B, \textit{The Cosmological constant
and dark energy}, 2003 \textit{Rev. Mod. Phys.} \textbf{75}, 559

\bibitem{fowlie} Fowlie A, Kowalska K, Roszkowski L, Sessolo E M and Tsai Y
L S, \textit{Dark matter and collider signatures of the MSSM}, 2013 \textit{%
Phys.Rev. D.} \textbf{88}, 055012

\bibitem{atlas} Aad G et al. (ATLAS Collaboration), \textit{Search for
Invisible Decays of a Higgs Boson Produced in Association with a Z Boson in
ATLAS}, 2014 \textit{Phys. Rev. Lett.} \textbf{112}, 201802

\bibitem{Frampton} Frampton P H, \textit{Angular Momentum of Dark Matter
Black Holes, }arXiv:1608.05009 [gr-qc]

\bibitem{Sasaki} Sasaki M, Suyama T, Tanaka T, Yokoyama S, \textit{%
Primordial Black Hole Scenario for the Gravitational-Wave Event GW150914},
2016 \textit{Phys. Rev. Lett. }\textbf{117}, 061101

\bibitem{millennium} Volker Springel, Simon D M White, Adrian Jenkins,
Carlos S Frenk, Naoki Yoshida, Liang Gao, Julio Navarro, Robert Thacker,
Darren Croton, John Helly, John A Peacock, Shaun Cole, Peter Thomas, Hugh
Couchman, August Evrard, Joerg Colberg and Frazer Pearce, \textit{Simulating
the joint evolution of quasars, galaxies and their large-scale distribution}%
, 2005 \textit{Nature} \textbf{435}, 629

\bibitem{nav} Navarro J F, Frenk C S and White S D M, \textit{The Structure
of cold dark matter halos}, 1996 \textit{ApJ} \textbf{462}, 563

\bibitem{valenzuela} Valenzuela O, Rhee G, Klypin A, Governato F, Stinson G,
Quinn T and Wadsley J , \textit{Is there Evidence for Flat Cores in the
Halos of Dwarf Galaxies?: The Case of NGC 3109 and NGC 6822}, 2006 \textit{%
ApJ} \textbf{657}, 773-789

\bibitem{jardel} Jardel J R, Gebhardt K, Fabricius M, Drory N and Williams M
J, \textit{Measuring Dark Matter Profiles Non-Parametrically in Dwarf
Spheroidals: An Application to Draco}, 2012 \textit{ApJ} \textbf{763}

\bibitem{burkert} Burkert A, \textit{Aspects of Dark Matter in Astro-and
Particle Physics}, 1997

\bibitem{teyssier} Teyssier R, Pontzen A, Dubois Y and Read J, \textit{%
Cusp-core transformations in dwarf galaxies: observational predictions},
2012 \textit{MNRAS} \textbf{429}, 3068

\bibitem{inoue} Inoue S and Saitoh T R, \textit{Shallowed cusp slope of dark
matter in disc galaxy formation through clump clusters}, 2011 \textit{MNRAS} 
\textbf{418}, 2527

\bibitem{klypin2009} Klypin A and Prada F, 2009 \textit{ApJ} \textbf{690},
1488

\bibitem{gt} Gergely L \'{A} and Tsujikawa S, \textit{Effective field theory
of modified gravity with two scalar fields: dark energy and dark matter},
2014 \textit{Phys. Rev. D} \textbf{89}, 064059

\bibitem{sin} Sin S J, \textit{Late time cosmological phase transition and
galactic halo as Bose liquid}, 1994 \textit{Phys. Rev. D} \textbf{50}, 3650

\bibitem{Sikivie1} {Sikivie} P, \textit{Caustic rings of dark matter}, 1998 \textit{Phys. Lett. B} \textbf{432}, 139-144

\bibitem{Sikivie2} {Sikivie} P, \textit{Caustic ring singularity}, 1999 \textit{Phys. Rev. D} \textbf{60}, 6

\bibitem{Gross1} Gross E P, \textit{Structure of a quantized vortex in boson
systems} \textit{Nuovo Cimento}, 1961 \textbf{20}, 454

\bibitem{Gross2} Gross E P, 1963, \textit{J. Math. Phys.} \textbf{4}, 195

\bibitem{Pitaevskii} Pitaevskii L P, \textit{Vortex Lines in an Imperfect Bose Gas}, 1961, \textit{Zh. Eksp. Teor. Fiz.} 
\textbf{40}, 646

\bibitem{rodr} Rodriguez-Montoya I, Magana J, Matos T and Perez-Lorenzana A, 
\textit{Ultra light bosonic dark matter and cosmic microwave background},
2010 \textit{ApJ} \textbf{721}, 1509

\bibitem{harko11} Harko T, \textit{Cosmological dynamics of dark matter
Bose-Einstein Condensation}, 2011 \textit{Phys.Rev.D} \textbf{83}, 123515

\bibitem{souza14} Souza J C C and Pires M O C, \textit{Discussion on the
energy content of the galactic dark matterBose-Einstein condensate halo in
the Thomas-Fermi approximation}, 2014 \textit{JCAP} \textbf{03}, 010

\bibitem{harko14} Harko T, \textit{Gravitational collapse of Bose-Einstein
condensate dark matter halos}, 2014, [arXiv:1403.3358]



\bibitem{lee} Lee J W, Lim S and Choi D, \textit{BEC dark matter can explain
collisions of galaxy clusters}, 2008 [arXiv:0805.3827v1]

\bibitem{velten} Velten H and Wamba E, \textit{Power spectrum for the
Bose-Einstein condensate dark matter}, 2012 \textit{Phys.Lett. B} \textbf{709%
} 1-5

\bibitem{boyan} Boyanovsky D, de Vega H J and Sanchez N, \textit{Constraints
on dark matter particles from theory, galaxy observations and N-body
simulations}, 2008 \textit{Phys. Rev. D} \textbf{77}, 043518

\bibitem{alma} Gonzalez-Morales A X, Diez-Tejedor A, Urena-Lopez L A and
Valenzuela O, \textit{Hints on halo evolution in SFDM models with galaxy
observations} 2012 \textit{Phys. Rev. D} \textbf{87} 02130

\bibitem{robl} Robles V H and Matos T, \textit{Flat Central Density Profile
and Constant DM Surface Density in Galaxies from Scalar Field Dark Matter},
2012 \textit{MNRAS} \textbf{422}, 282-289

\bibitem{dwor} Dwornik M, Keresztes Z and Gergely L \'{A}, \textit{Recent
Development in Dark Matter Research} 2014 \textit{Nova Science Publishers}
p. 195-219

\bibitem{pita} Pitaevskii L P and Stringari S, \textit{Bose-Einstein
Condensation}, 2003 \textit{Oxford University Press Inc., New York.}

\bibitem{GrossmannHolthaus} Grossmann S and Holthaus M, 1995 \textit{Phys. Lett. A} \textbf{208}, 188

\bibitem{KetterleDruten} Ketterle W and van Druten N J, \textit{Two-Step
Condensation of the Ideal Bose Gas in Highly Anisotropic Traps}, 1996 
\textit{Phys. Rev. A} \textbf{54}, 656

\bibitem{KristenToms} Kristen K and Toms D J, \textit{Bose-Einstein
condensation of atomic gases in a general harmonic-oscillator confining
potential trap}, 1996 \textit{Phys. Rev A} \textbf{54}, 4188

\bibitem{Haugerudetal} Haugerud H, Haugset T and Ravndal F, \textit{%
Bose-Einstein condensation in anisotropic harmonic traps }, 1997 \textit{%
Phys.Lett. A} \textbf{225}, 18

\bibitem{Giorginietal} Giorgini S, Pitaevskii L and Stringari S, \textit{%
Theory of ultracold atomic Fermi gases}, 1996 \textit{Phys. Rev. A} \textbf{%
54}, R4633

\bibitem{Glaumetal} Glaum K, Pelster A, Kleinert H and Pfau T, \textit{%
Critical Temperature of Weakly Interacting Dipolar Condensates}, 2007 
\textit{Phys. Rev. Lett.} \textbf{98}, 080407

\bibitem{Schutte} Sch\"{u}tte M and Pelster A, \textit{Critical Temperature
of a Bose-Einstein Condensate with 1/r Interactions}, 2008 Proceedings of
the 9th International Conference, 23-28, September, 2007, Dresden, Germany,
Eds. Janke W. and Pelster A., World Scientific Publishing Co. Pte. Ltd.,
2008. ISBN \#9789812837271, pp. 417-420

\bibitem{dalfovo} Dalfovo F, Giorgini S, Pitaevskii L P and Stringari S, 
\textit{Theory of Bose-Einstein condensation in trapped gases}, 1999 \textit{%
Rev. Mod. Phys.} \textbf{71}, 463

\bibitem{anderson} Anderson M H, Ensher J R, Matthews M R, Wieman C E and
Cornell E A, \textit{Observation of Bose-Einstein Condensation in a Dilute
Atomic Vapor}, 1995 \textit{Science} \textbf{269}, 198

\bibitem{Han98} Han D J, Wynar R H, Courteille Ph. and Heinzen D J, \textit{%
Bose-Einstein Condensation of Large Numbers of Atoms in a Magnetic
Time-Averaged Orbiting Potential Trap}, 1998 \textit{Phys. Rev. A} \textbf{57%
}, R4114

\bibitem{Ernst98} Ernst U, Marte A, Schreck F, Schuster J and Rempe G, 
\textit{Bose-Einstein Condensation in a Pure Ioffe-Pritchard Field
Configuration}, 1998 \textit{Europhys. Lett.} \textbf{41}, 1

\bibitem{Davis95} Davis K B, Mewes M O, Andrews M R, van Druten N J, Durfee
D S, Kurn D M and Ketterle W, \textit{Bose-Einstein Condensation in a Pure
Ioffe-Pritchard Field Configuration}, 1995 \textit{Phys. Rev. Lett.} \textbf{%
\ 75}, 3969

\bibitem{Hau98} Hau L V, Busch B D, Liu C, Dutton Z, Burns M M and
Golovchenko J A, \textit{Near Resonant Spatial Images of Confined Bose--%
Einstein Condensates in the 4-Dee Magnetic Bottle}, 1998 \textit{Phys. Rev. A%
} \textbf{58}, R54

\bibitem{Bradley95} Bradley C C, Sackett C A, Tollett J J and Hulet R G, 
\textit{Evidence of Bose-Einstein condensation in an atomic gas with
attractive interactions}, 1995 \textit{Phys. Rev. Lett.} \textbf{75}, 1687

\bibitem{Madelung} Madelung E., \textit{Quantum theory in hydrodynamic form}, 1926, \textit{Zeitschrift f\"{u}r Physik} 
\textbf{38}, 322

\bibitem{Sonego} Sonego S, \textit{Interpretation of the hydrodynamical
formalism of quantum mechanics}, 1991 \textit{Found. Phys.} \textbf{21}, 1135

\bibitem{Wang} Wang X Z, \textit{Cold bose stars: Selfgravitating
Bose-Einstein condensates}, 2001 \textit{Phys. Rev D} \textbf{64}, 124009

\bibitem{Liebetal} Lieb E H, Seiringer R and Yngvason Y, \textit{A rigorous
derivation of the Gross-Pitaevskii energy functional}, 2000 \textit{Phys.
Rev. A} \textbf{61}, 043602

\bibitem{ser} S\'{e}rsic J L, 1968, \textit{Atlas de Galaxias Australes},
Cordoba, Argentina, Observatorio Astronomico

\bibitem{free} Freeman K C, \textit{On the disks of spiral and SO Galaxies},
1970 \textit{ApJ} \textbf{160}, 811

\bibitem{palu} Palunas P and Williams T B, \textit{Maximum Disk Mass Models
for Spiral Galaxies}, 2000 \textit{ApJ} \textbf{120}, 2884

\bibitem{imp} Impey C and Bothun G, \textit{Low Surface Brightness Galaxies}%
, 1997 \textit{ARA\&A} \textbf{35}, 267

\bibitem{mcg} McGaugh S S, \textit{Oxygen abundances in low surface
brightness disk galaxies}, 1994 \textit{ApJ} \textbf{426}, 135

\bibitem{neil} O'Neil K, Bothun G D, Schombert J, Cornell M E and Impey C D, 
\textit{A Wide Field CCD Survey for Low Surface Brightness Galaxies.II.Color
Distributions, Stellar Populations, and Missing Baryons}, 1997 \textit{ApJ} 
\textbf{144}, 244

\bibitem{beij} Beijersbergen M, de Blok W J G and van der Hulst J M, \textit{%
Surface photometry of bulge dominated low surface brightness galaxies}, 1999 
\textit{A\&A} \textbf{351}, 903

\bibitem{blok} de Blok W J G and Bosma A, \textit{High-resolution rotation
curves of low surface brightness galaxies}, 2002 \textit{A\&A} \textbf{385},
816

\bibitem{karac} Karachentsev I D, Karachentseva V E, Huchtmeier W K and
Makarov D I, \textit{A Catalog of Neighboring Galaxies }, 2004 \textit{ApJ} 
\textbf{127}, 2031

\bibitem{tamm} Tammann G A, \textit{Dwarf Galaxies in the Past, in Dwarf
Galaxies}, 1994 \textit{ESO Conference and Workshop Proc No. 49} p. 3

\bibitem{matt} Matthews L D and Gallagher J S, \textit{B and V CCD
photometry of southern, extreme late-type spiral galaxies}, 1997 \textit{ApJ}
\textbf{114}, 5

\bibitem{mate} Mateo M, \textit{Dwarf galaxies of the Local Group }, 1998 
\textit{ARA\&A} \textbf{36}, 435

\bibitem{capo} Capozziello S, Cardone V F and Troisi A, \textit{Low surface
brightness galaxy rotation curves in the low energy limit of Rn gravity: no
need for dark matter?}, 2007 \textit{MNRAS} \textbf{375}, 1423

\bibitem{Helou1991}{Helou} G, {Madore} B F, {Schmitz} M, {Bicay} M D, {Wu} X, and {Bennett} J, \textit{The NASA/IPAC extragalactic database.}, 1991 \textit{ASSL} \textbf{171}, {89-106}

\bibitem{McGaugh2000} {McGaugh} S S, {Schombert} J M, {Bothun} G D, and {de Blok} W J G,\textit{The Baryonic Tully-Fisher Relation}, 2000 \textit{ApJL} \textbf{533}, L99-L102
	
\bibitem{McGaugh2010} {McGaugh} S S and {Wolf} J, \textit{Local Group Dwarf Spheroidals: Correlated Deviations from the Baryonic Tully-Fisher Relation}, 2010 \textit{ApJ} \textbf{722}, 248-261

\bibitem{pers} Persic M, Salucci P and Stel F, \textit{The universal
rotation curve of spiral galaxies---I. The dark matter connection}, 1996 
\textit{MNRAS} \textbf{281}, 27

\bibitem{PecceiQuinn} Peccei R D and Quin H, \textit{CP Conservation in the Presence of Pseudoparticles}, 1977, \textit{Phys. Rev. Lett.} \textbf{38}, 1440

\bibitem{axionL1} Asztalos S J, Caosi G, Hagmann C et al., \textit{The Axion Dark Matter eXperiment}, 2011, XXXI Physics in Collision, Vancouver, BC Canada, August 28 - September 1, arXiv:1112.1167

\bibitem{axionL2} Rosenberg L J, \textit{Dark-matter QCD-axion searches} 2013, \textit{PNAS} \textbf{112}, 12278

\bibitem{pire} Pires M O C and de Souza J C C, \textit{Galactic cold dark
matter as a Bose-Einstein condensate of WISPs}, 2012 \textit{JCAP} \textbf{11}, 024
\end{thebibliography}
\end{document}